\documentclass[fleqn,twoside]{article}%
\usepackage{amsmath}
\usepackage{amsfonts}
\usepackage{amssymb}
\usepackage{graphicx}
\usepackage{bm}
\usepackage{xcolor}
\usepackage{cite}
\def\be{\begin{equation}}
\def\ee{\end{equation}}

\begin{document}
\title{Issue of Branched Hamiltonian, Inflation and indispensability of scalar
field in generalized Teleparallel theories.}
\author{Dalia Saha$^1$, Jyoti Prasad Saha$^2$, Abhik kumar sanyal$^{3*}$}
\maketitle
\noindent
\begin{center}
\noindent
$^{1}$ Dept. of Physics, Jangipur College, Murshidabad, West Bengal, India - 742213, \& Dept. of Physics, University of Kalyani, Nadia, West Bengal, India - 741235.\\
$^{2}$ Dept. of Physics, University of Kalyani, Nadia, West Bengal, India - 741235.\\
$^{3*}$ Dept. of Physics, Jangipur College, Murshidabad, West Bengal, India - 742213, \& Calcutta
Institute of Theoretical Physics, Bignan Kutir, 4/1, Mohanbagan Lane, Kolkata,India - 700004.\\
\end{center}
\footnotetext[1]{
\noindent Electronic address:\\
$^1$daliasahamandal1983@gmail.com\\
$^2$jyotiprasadsaha@gmail.com\\
$^{3*}$sanyal\_ak@yahoo.com\\}

\begin{abstract}
Generalized Teleparallel gravity theories were proposed as alternatives to the dark energy and modified theories of gravity. However, minimal generalization of TEGR and STEGR lead to the issue of branched Hamiltonian. Here, in the FLRW background we fix the coupling parameters of the said model in view of energy condition and show that the theory is free from the issue of Branched Hamiltonian unlike second-order Lanczos-Lovelock gravity. We also study scalar field driven inflation, which shows excellent agreement with recent observations. Nonetheless, a viable radiation era requires the very presence of the scalar field, which is retained at the late universe as dark energy. Thus the fundamental claim to act as an alternative to the dark energy is dubious.
\end{abstract}
\textbf{keywords:} Teleparallel gravity, Energy conditions, Branched Hamiltonian, Cosmological evolution.\\

\section{Introduction}\label{sec1}
In recent years, in analogy to the modified theories of gravity, such as $f(R)$ \cite{0a,0a.1}, $f(R, \mathcal G)$ \cite{0b,0c,0d,0e} etc., where $R$ and $\mathcal G$ are the Ricci scalar and Gauss-Bonnet scalar respectively, an altogether different formulation of the theory of gravity, known as `Generalized Teleparallel Theory of Gravity' has drawn a lot of attention. General Theory of Relativity (GTR) based on Levi-Civita connection ascribes gravity to the space-time curvature. Despite its overall success, Einstein himself later made an alternative prescription where gravity appears as a consequence of torsion, which is called teleparallel gravity due to the absence of Riemann curvature tensor. This proposal was made in an attempt to unify electromagnetism with gravity, which was a failure though \cite{1a}. Afterwards the theory rejuvenated in an attempt to solve the gravitational energy problem \cite{1b}, to formulate gravity as a gauge theory \cite{1c,1d}, and to improve the action principle \cite{1e} etc. Such a theory of gravity is characterized by the torsion tensor $({\mathrm{T}_{\mu\nu}})$ instead of the Ricci curvature tensor $(R_{\mu\nu})$ and one assembles the so-called torsion scalar ${\mathrm{T}}$ constructed from the torsion tensor $\mathrm{T}_{\mu\nu}$, which plays the role of the scalar curvature $R$ in the Einstein-Hilbert action. However, this theory differs from GTR by a total derivative term and therefore is essentially dynamically equivalent to GTR, which is dubbed as `Teleparllel Equivalent of General Relativity' or TEGR under abbreviation. In the recent years a generalized version $f(\mathrm T)$ of the theory resurrected to act as alternative to the dark energy and modified theories of gravity. While curvature is non-vanishing along with the torsion in Einstein-Cartan theory and also in Palatini formalism formulated in the metric affine environment, teleparallel theory is based on general affine connection in a curvature free environment. The associated affine connection in teleparallel theory is a combination of the tetrads (originally called vielbeins which define local reference frame) and the spin connections. Being a gauge theory, it is possible to choose different gauges resulting in different sets of connections and consequently different field equations. However, the theory is simplified in the pure tetrad, also called the good tetrad (Use of a good tetrad simplifies the field equations considerably and satisfies them non-trivially, without restricting parameters of the theory. Bad tetrads in contrast, yield solutions mostly in GTR limit), for example in the Weitzenb\"ock gauge where the spin connection is set to vanish \cite{2,2a}. Due to considerable simplification of the field equations, it plays a major role in the study of cosmology. We dub it as `Generalized Metric Teleparallel theory of Gravity' (GMTG), where the action contains an arbitrary function of torsion scalar $f(\mathrm{T})$ \cite{2}. There exists yet another teleparallel theory, which describes gravity from the affine connection with vanishing curvature and torsion, allowing its non-metricity tensor $(Q_{\lambda\mu\nu} = \nabla_\lambda g_{\mu\nu} \ne 0)$ to be responsible for gravitational interaction. Here, $\nabla$ is the covariant derivative satisfying curvature-free and torsion-free conditions, while $g_{\mu\nu}$ is the metric tensor. This particular theory is dubbed as `symmetric teleparallel theory of gravity', since the tensor $Q_{\lambda\mu\nu}$ is symmetric in the last two indices. One can construct the non-metricity scalar $Q$ from a combination of third rank non-metricity tensor $Q_{\lambda\mu\nu}$ and its vector traces $Q_\lambda$ and $\hat Q_\lambda$, which plays the role of $R$. This theory is also dynamically equivalent to GTR and is called STEGR (Symmetric Teleparallel Equivalent of General Relativity). Nonetheless, in accordance to the modified $f(R)$ theory of gravity and GMTG $f(\mathrm{T})$, the `Generalized Symmetric Teleparallel Gravity' (GSTG) has been proposed and formulated, in which the non-metriciticy scalar is replaced by an arbitrary function $f(Q)$. These teleparallel theories not only are said to combat the cosmic puzzle \cite{3,4} but also exhibit many features which are absent from GTR and therefore are richer in structure than GTR. Being formed in the curvature-free environment, these theories are called `Alternative theories of gravity', while generalized versions of GTR are known as modified theories of gravity.\\

The main advantage of generalized teleparallel theories is that these theories result in second order field equations rather than the fourth or even higher order ones, as for modified $f(R)$ gravity theories. Consequently, `Ostrgradsky's instability' is avoided \cite{5} straight away, apparently though. Let us mention that in the case of `modified theory of gravity' too, unless $f(R)$ is chosen judiciously, such that $f(R)_{,R} > 0$ and $ f(R)_{,RR} > 0$, where comma denotes derivative, the theory suffers from the presence of ghost degrees of freedom\footnote{The modified theory contains extra degree of freedom, for example a scalar field. If this field acquires a kinetic term with opposite (wrong) kinetic term, i.e., ${1\over 2}g_{\mu\nu}\phi^{,\mu}\phi^{,\nu} \rightarrow -{1\over 2}g_{\mu\nu}\phi^{,\mu}\phi^{,\nu}$, ghost appears. As a result, the Hamiltonian is not bounded from below and the system continuously acquires lower energy states, driving infinite instability. Specifically, the presence of ghosts results in runaway decay of physical systems, Thus, the background state grows exponentially over time, destroying cosmological structures such as stars, galaxies and clusters. In the quantum domain unitarity is lost, while positive and negative energy particles are created spontaneously with uncontrolled decay of space-time, which destroys the background space-time.} and stability criterion. Further, as mentioned, both these teleparallel theories are claimed to be able to combat with the cosmic puzzle, viz., early decelerated expansion followed by the currently observed accelerated expansion of the universe, without seeking the presence of scalar fields \cite{6.1,7.1,8.1,9.1,10.1,11.1,12.1,Bam}. Additionally, static spherically symmetric solutions have also been explored \cite{Marco1, Marco2}. These are particularly the primary reasons for intense interest and substantial amount of research in `Generalized Teleparallel Gravity' theories in recent years. \\

Unfortunately, GMTG is afflicted with a glitch. The trendy one (Weintzenb\"ock gauge), being constructed out of vanishing spin connection, is not Locally Lorentz Invariant (LLI) and possesses additional spurious degrees of freedom \cite{LLI1,LLI2}, unlike $f(R)$ theories. To be specific, $f(\mathrm{T})$ theories give rise to $16$ equations \cite{fT/lli} instead of $10$ equations of GTR. Further, $f(\mathrm{T})$ connections obey the conditions $Q_{\lambda\mu\nu} = 0$, which are $40$ independent equations because of the symmetry of non-metricity tensor $Q_{\lambda\mu\nu}$ in the last two indices. Later, treating both the metric and the spin connection as independent variables, the issue of LLI is alleviated \cite{Tcov1,Tcov2}, but the theory becomes difficult to handle in general. Further, $f(\mathrm{T})$ gravity is afflicted by the presence of ghost degrees of freedom and suffers from strong coupling issues\footnote{If the kinetic term associated with the extra degree of freedom is extremely small, then interaction diverge at arbitrarily low energies and perturbation breaks down. In a nutshell, dominance of non-linear interactions at arbitrarily low scales indicate strong coupling.}, which put stronger limitations to the $f(\mathrm{T})$ connections \cite{Tpert,fT1,fT2}. In the mean time, GSTG came into the limelight, in which all these issues initially was thought to be absent from $f(Q)$ theory. Because $f(Q)$ connections abide by the vanishing torsion tensor criteria, so $\mu = \nu$ case is trivially satisfied in the relation ${\mathrm{T}}^\gamma_{\,\,\,\mu\nu} = 0$, and only $24$ independent equations emerge. Nonetheless, the subject of counting the number of degrees of freedom is disputed and arguable. For further detailed understanding of these interesting issues, we refer to \cite{fQfT, fQfT1, fQfT2, fQfT3, de/epjc} and the references therein.\\

In the non-metricity-GSTG theory, field equations are often simplified under the choice of a specific gauge, called the `coincidence gauge' \cite{coincident, lin, cosmography, de/prd, signa, redshift, perturb, dynamical1, sf1, sf2}. In `coincidence gauge', all the connections vanish making calculation simpler by reducing the `covariant derivative' to merely `partial derivative' $(Q_{\lambda\mu\nu} = \nabla_\lambda g_{\mu\nu} = \partial_\lambda g_{\mu\nu}\ne 0)$. Lamentably, this is again a non-LLI theory and recent study following perturbative analysis reveals that GSTG theory in the background of maximally symmetric spatially flat space also suffers from severe pathologies such as `strong coupling issues' and `Ostrogradski's instability' due to the presence of `ghost degrees of freedom' \cite{fT2, p1, p2, p3}. But even more serious problem is that the theory is not diffeomorphic invariant in general\footnote{A theory is invariant under diffeomorphism if the field equations remain unchanged as the space-time is smoothly deformed or reparametrized. Under diffeomorphism invariance, the laws of physics remain independent of arbitrary transformation. In $f(Q)$ theory, it is required to consider a set of eleven variables $(N, N_i, h_{ij},\Phi)$ where the shift vector $(Ni)$ and the auxiliary variable $\Phi$ are found to be the dynamical but non-propagating variables carrying $1/2$ degree of freedom each. The momentum associated with the auxiliary variable $\Phi$ appears as the first derivative of the shift vector $p_\Phi = {\sqrt{h}\over N} f_{\phi\phi}(\partial_i N_i)$ which spoils diffeomorphism invariance.} \cite{H1,H2,H3}. Since the shift vector ($N_i$) is the root of trouble, so in its absence, such as for the Robertson-Walker (RW) and the anisotropic Bianchi metrics the pathology disappears. In fact it has been procured in the RW space-time \cite{DA}.\\

The Local Lorentz Invariance (LLI) of $f(Q)$ gravity is assured once coincidence gauge is withdrawn and one considers the affine connection to be an independent variable in addition to the metric \cite{coincident, Qcov1}, as in the case of GMTG. In this attempt, it has been found that the RW space-time admits four different non-trivial (non-coincidence gauge) connections, out of which three are for the spatially flat ($k = 0$) space-time and one for non-flat ($k =\pm 1$) space-time \cite{NCG1,NCG2}. However, the second and the third connections have been found to suffer from the stability issue and might also from singularity issues \cite{2.3}. In a recent study it has been further revealed that these two connections lead to $f_{,QQ} = 0$ and so cannot even be generalized to $f(Q)$ gravity. Thus, these are therefore simply STEGR, which are dynamically equivalent to GTR \cite{DA2}. On the contrary, the very first ($k = 0$) and also the fourth (supposedly for $k = \pm 1$, but found to allow only $k = -1$ for real parameteric values) connections may be generalized and also exhibit diffeomorphic invariance (in the absence of shift vector), though the Hamiltonians are apparently non-tractable \cite{DA2}. In any case, the very presence of diffeomorphism invariance suggests that the theory is still open for further scrutiny at least for the first and the fourth connections. In this manuscript we shall deal with spatially flat case and so appraise the first connection only, leaving out the fourth connection.\\

Above discussions reveals that generalised teleparallel gravity theories are afflicted with several irremediable issues and therefore not as promising as was thought earlier. Our current intention is to explore if at all these theories have merit. For this purpose we consider spatially flat $(k = 0)$ spherically symmetric Robertson-Walker (RW) metric in the following form,
\begin{eqnarray}\label{RW} ds^2 = -dt^2 + a^2(t)\left[dr^2 + r^2(d\theta^2 + r^2 sin^2 \theta d\phi^2)\right],\end{eqnarray}
where $a(t)$ is the scale factor. One interesting fact regarding symmetric teleparallel theory is that the coincidence gauge leads to the same form of $Q = - 6 H^2$, ($H$ being the Hubble parameter) along with identical field equations as for the very first non-trivial (non-coincidence gauge) connection in the flat FLRW background. Also dynamics remains unaltered since the connection variation equation is trivially satisfied. Firstly, we discuss the issue of the Branched Hamiltonian, which was earlier dealt with GMTG theory \cite{IJJ}, under the natural choice $f(\mathrm{T}) = \alpha \mathrm{T} + \beta \mathrm{T}^2$ with proper tetrad fields. It may be noted that, the torsion scalar also reads as $\mathrm{T}= -6H^2$ and it was pointed out that the cosmological field equations in flat FLRW background coincide in $f(\mathrm{T})$ and $f(Q)$ under said gauge choices \cite{Qcov1}. Therefore, the issue of Branched Hamiltonian holds for the natural choice $f(Q) = \alpha Q + \beta Q^2$ as well. Thus, we treat both of them on the same footing, using the notation $\mathrm{T} = Q = X$ and exhibit the fact that the energy condition restricts the sign of the coupling parameters to $\alpha > 0, \beta < 0$, for which the pathology of branching disappears. This is one virtue of the theory in comparison with well-versed Lanczos-LoveLock theory, whose field equations are also second order. Next, since pure vacuum solution doesn't admit arbitrary form of $f(X)$ required to drive inflation, so we study scalar field driven inflation for four different potentials. It is found that the parameters of the theory for the two flat potentials can be reasonably constrained so that the inflationary parameters lie within the latest published observational data received from Planck and ACT \cite{Planck1,Planck2,Planck3,ACT1,ACT2,ACT3} and also admit graceful exit as well. This is yet another virtue. Nonetheless, we further exhibit the fact that if the scalar field is completely used up during particle creation (reheating), then a viable Friedmann-like decelerated radiation era is not achievable. A very little presence of the scalar field although serves the purpose, it escalates by a huge amount and remains constant in the matter-dominated (pressure-less dust) era. This is the vice, since the fundamental inspiration of building generalized teleparallel theories as alternative to the dark-energy issue goes astray in isotropic and homogeneous space-time, while dark energy has not yet been detected.\\

The structure of the paper is as follows. A brief discussion of geometry related with both the teleparallel theories of gravity is provided in the section that follows. In section 3, in the FLRW background we restate the early vacuum-dominated era to perceive the reason for choosing a viable form of $f(X)$ by hand and the need for a scalar field driven inflation. We choose $f(X) = \alpha_1 X + \beta_1 X^2$, which is a minimal generalization over GTR and summarize the issue of Branched Hamiltonian studied earlier \cite{IJJ} in section 4. We then fix the sign of the coupling constants ($\alpha_1 > 0, \beta_1 < 0$) in view of the energy conditions and elucidate the fact that the said pathology does not actually appear. Next, we study a scalar-driven slow-roll inflation in the same section taking into account four different scalar field potentials and formulate reheating for the flat exponential potential. In section 5, we show that a viable radiation dominated era is only admissible in the presence of a scalar field, however small. Lamentably, the scalar escalates by a huge amount and thereafter remains constant in the pressure-less dust dominated era, exhibiting Friedmann-like decelerated expansion. Finally, in section 6, we briefly highlight the covariant scalar-tensor formalism of $f(\mathrm{T})$ and the scalar-vector-tensor formalism of $f(Q)$ theories which do not require the presence of any additional scalar field, but do not procure new result. Section 7 concludes our work.

\section{Geometry of teleparallel gravity theories.}
While the Riemann curvature tensor $({R^\alpha}_{\beta\mu\nu})$ based on Levi-Civita connection is the building block of GTR, the general affine connection may be decomposed self-consistently into a combination of the Levi-Civita connection (components of this connection in a coordinate basis are called the Christoffel symbols), the Contortion tensor and the Disformation tensor \cite{2a}, which is given by the following expression,
\begin{eqnarray}\label{2.1} {\Gamma^\alpha}_{\mu\nu} = \{_\mu{^\alpha}_\nu\}  + {K^\alpha}_{\mu\nu} + {L^\alpha}_{\mu\nu},\end{eqnarray}
where, the Levi-Civita connection, the contortion tensor and the disformation tensor are expressed as,
\begin{eqnarray}\label{2.2} \begin{split}& \{_\mu{^\alpha}_\nu\} = \frac{1}{2} g^{\alpha\lambda}\left(g_{\mu\lambda,\nu}+g_{\nu\lambda,\mu}-g_{\mu\nu,\lambda}\right),\\&
{K^\alpha}_{\mu\nu} = \frac{1}{2}g^{\alpha\lambda}\left(\mathrm {T}_{\lambda\mu\nu}-\mathrm{T}_{\mu\lambda\nu}-\mathrm{T}_{\nu\lambda\mu}\right),\\&{L^\alpha}_{\mu\nu} = \frac{1}{2}g^{\alpha\lambda}\left(Q_{\lambda\mu\nu}-Q_{\mu\lambda\nu}-Q_{\nu\lambda\mu}\right),\end{split}\end{eqnarray}
The Levi-Civita connection is formed out of the first derivatives of the metric tensor $(g_{\mu\nu})$, while the contortion and the disformation tensors are formed out of the linear combinations of the torsion tensor $(\mathrm{T}_{\alpha\mu\nu})$ and the non-metricity tensor $(Q_{\alpha\mu\nu})$ respectively. The general Riemann curvature tensor is formed out of the combination of the general affine connection and its derivatives as,
\begin{eqnarray} \label{Curv} {R^\alpha}_{\beta\mu\nu} = \partial_\mu{\Gamma^\alpha}_{\nu\beta} - \partial_\nu{\Gamma^\alpha}_{\mu\beta} + {\Gamma^\alpha}_{\mu\sigma}{\Gamma^\sigma}_{\nu\beta} - {\Gamma^\alpha}_{\nu\sigma}{\Gamma^\sigma}_{\mu\beta}.\end{eqnarray}
If the connection is not symmetric, then in the flat space, i.e., for vanishing Riemann curvature, the above curvature tensor may be expressed solely out of the contortion tensor, which we denote as ${\bar R^\alpha}_{~\beta\mu\nu}$. On the contrary, if the connection is symmetric, it \eqref{Curv} may be expressed in terms of the sum of the Riemann curvature tensor formed out of Levi-Civita connection and a combination of the disformation tensors and we denote it by ${\tilde R^\alpha}_{~\beta\mu\nu}$. In the next subsection, we present an outline regarding the geometries of the two teleparallel theories.

\subsection{Metric teleparallel theory of gravity:}
The general torsional teleparallel connection contains both the tetrads (vielbeins) and the spin connections and is expressed as, 
\be \label{Gm}{\mathrm {\Gamma}^\alpha}_{\mu\nu} = {e^\alpha}_a\partial_\mu{e^a}_\nu + {e^\alpha}_a{\omega^a}_{b\mu}{e^b}_\nu,\ee
where, ${e^\mu}_a$ are the tetrad fields and ${\omega^{ab}}_\mu$ are the spinor fields. In $f(\mathrm{T})$ theory equation of motion are derived by varying the action with respect to the tetrad field and the spin connection. While, the spin connection in GTR essentially is the Levi-Civita connections in the tetrad basis (${\omega^{ab}}_\mu|_{LC} = {e^a}_\nu[\partial_\mu e^{b\mu} + \{_\mu{^\nu}_\rho\} e^{b\rho}]$) and are dynamical variables, the same in torsional gravity is given as ${\omega^a}_{b\mu}|_{TG} = {\Lambda^a}_c\partial_\mu({\Lambda^{-1})^c}_b$, which is flat (zero curvature) and ensures LLI\footnote{If at each space-time point one can use a local inertial frame using tetrads ${e^a}_\mu$, relating space-time indices (greek) to local Lorentz indices (latin) then the theory is LLI. Note that, in the absence of spin connection $\Gamma$ \eqref{Gm} appears as partial derivative and under position dependent Lorentz transformation (${e^a}_\mu \rightarrow {\Lambda^a}_b(x) {e^b}_\mu$), $\partial_\mu {e^a}_\nu$ does not transform covariantly, since derivative also act on the boost. As a result the torsion scalar $\mathrm T$ is also not invariant under local Lorentz transformations of the tetrad. Introducing spin connection (${\omega^a}_{b\mu}$), which acts only as a gauge field, partial derivative is replaced by covariant derivative and reads as $D_\mu {e^a}_\nu = \partial _\mu {e^a}_\nu + {\omega^a}_{b\mu}{e^b}_\nu - {\Gamma^\rho}_{\nu\mu}{e^a}_\rho$ which transforms covariantly. In essence, the geometric terms of the spin connection cancel the non-inertial pseudo terms. Thus, covariance under LLT requires simultaneous transformation of tetrad and the spin connection.}. Nonetheless, since spin connection in $f(\mathrm T)$ is constrained to be flat, therefore it contains only inertial effect (and is purely gauge) analogous to fictitious forces appearing in accelerated frame of reference. Under variation, the spin connection only imposes consistency and LLI without affecting dynamics and additional dynamical degree of freedom. Of-course the presence of spin connection makes $f(\mathrm{T})$ theory difficult to handle in general, since finding the correct spin connection for a complicated tetrad can be quite complicated (but not in flat FLRW background). Let us mention that the situation is drastically simplified in its absence, because in any $f(\mathrm{T})$ theory, it is always possible to find a tetrad field in which the spin connection coefficients vanish, regardless of the gauge choice. The only role played by the gauge is to establish the tetrad field in which the spin connection coefficients vanish. It may be mentioned that the Weitzenb\"ock connection vanishes in a particular tetrad, or in a particular gauge. The flat RW metric \eqref{RW} may be expressed in Cartesian coordinate system as,
\be\label{RWC} ds^2 = -dt^2 + a(t)^2 [dx^2 + dy^2 + dz^2]\ee
which admits the diagonal Cartesian tetrad, which are called good or proper tetrad. In such a situation spin connection may be set to vanish as in the case of Weitzenb\"ock gauge and only the tetrad fields act as dynamical variables, being the fundamental component of metric teleparalallism. The tetrad and the co-tetrad fields ${e^\mu}_a$ and ${e^a}_\mu$ respectively are restricted by the following relations,
\begin{eqnarray} \label{2.3} {e^\mu}_a {e^a}_\nu = {\delta^\mu}_\nu, \;\;{e^\mu}_a{e^b}_\mu = {\delta^b}_a,\end{eqnarray}
while the `metric tensor' is expressed as,
\begin{eqnarray} \label{2.4} g_{\mu\nu} = \eta_{ab}{e^a}_\mu {e^b}_\nu,\end{eqnarray}
where, $\eta_{ab} = \{-1,+1,+1,+1\}$ is the tetrad metric, which is simply the Minkowskian metric in the tangent space and the tetrad one-forms are 
\be e^a = {e^a}_\mu dx^\mu, ~{e^a}_\mu = \mathrm{diag}\big(1, a(t), a(t), a(t)\big).\ee
In the above, the Greek indices $\mu$ and $\nu$ are the space-time coordinate indices on the manifold, while the Latin indices $a$ and $b$ label the same for the tangent space and both run over ($0-3$). As already mentioned, $f(\mathrm{T})$  theory constructed in the absence of the spin connection simplifies the geometry to a large extent by utilizing the constraint ${\bar R^\alpha}_{~\beta\mu\nu}  = 0$ \eqref{Curv}, where the over-bar stands for the Riemann curvature tensor formed out of the so-called Weintzenb\"ock gauge \cite{fQfT}. The simplified connection therefore reads as
\begin{eqnarray}\label{2.5} {\Gamma^\alpha}_{\mu\nu} = {e^\alpha}_a \partial_\nu {e^a}_\mu,\end{eqnarray}
so that,
\begin{eqnarray}\label{2.6} \begin{split}& {\mathrm{T}^{\alpha}}_{\mu\nu} = \left({\Gamma^\alpha}_{\mu\nu}-{\Gamma^\alpha}_{\nu\mu}\right) = {e^{\alpha}}_{a}[\partial_{\mu}{e^{a}}_{\nu}-\partial_{\nu}{e^{a}}_{\mu}] ,\\&
{K^{\mu\nu}}_{\alpha} \equiv  {1\over 2}\left[{{\mathrm{T}^{\mu}}_{\alpha}}^\nu+{{\mathrm{T}^{\nu}}_{\alpha}}^\mu-{\mathrm{T}_{\alpha}}^{\mu\nu}\right].\end{split}\end{eqnarray}
A tensors ${S^{\mu\nu}}_\alpha$ called the super-potential, expressed as,
\begin{eqnarray}\label{2.7}{S^{\mu\nu}}_\alpha =  {1\over 2}\left({K^{\mu\nu}}_\alpha +  {\delta^\mu}_\alpha{\mathrm{T}^{\beta\nu}}_\beta - {\delta^\nu}_\alpha{\mathrm{T}^{\beta\mu}}_\beta\right),\end{eqnarray}
is used to construct the quadratic torsion scalar
\begin{eqnarray}\label{2.8} \mathrm{T} = {S_\alpha}^{\mu\nu}{\mathrm{T}^\alpha}_{\mu\nu} = -6H^2.\end{eqnarray}
But as already mentioned, metric teleparallel gravity with spin zero proper tetrad such as Wietzenb\"ock connection is not LLI and therefore it is required to associate the spin connection also\footnote{Note that field equations depend on both the choice of tetrad and the spin connection. So wrong choice of either leads to different and wrong field equations. While flat FLRW metric has maximal spatial symmetry (homogeneity and isotropy), the tetrad is almost uniquely fixed by symmetry. This is not true general, as one requires to recognize which part of the tetrad is inertial and which part is gravitational and the corresponding inertial spin connection has to be reconstructed. In covariant teleparallel gravity the reference tetrad must satisfy ${T^a}_{\mu\nu}(h_r, \omega) = 0$ which is non-trivial in general.}. In order to include spin connection, one requires to go over to the spherical coordinate system of the said metric \eqref{RW}. As a result, the tetrad fields are now given by,
\be\label{tetrad} {e^a}_\mu = \mathrm{diag}\big(1,a(t), a(t) r, a(t) r sin\theta\big).\ee
However, a diagonal spherical tetrad differs from a diagonal Cartesian tetrad by a local rotation $({\Lambda^a}_b (x))$ that depends both on $\theta$ and $\phi$. Under such local Lorentz transformation, the tetrad derivative picks up extra $\partial_\mu \Lambda (x)$ pieces while the torsion scalar contains extra terms and turns out inhomogeneous. Nonetheless, if a flat (purely inertial) spin connection ${\omega^a}_{b\mu}$ is associated whose non-vanishing coefficients are,
\be\label{SC} {\omega^1}_{2\theta} = -1,~{\omega^1}_{3\theta} = -sin\theta,~{\omega^2}_{3\phi} = -cos\theta,~\omega_{ab\mu} = - \omega_{ba\mu}\ee
which correspond to rotations in $(r,\theta)$ plane, $(r,\phi)$ plane and $(\theta,\phi)$ plane respectively, then the said spatial derivatives $\partial_\mu \Lambda (x)$ are cancelled. Hence, in the covariant formulation, the spin connection ${\omega^a}_{b\mu}$ encodes purely inertial effects and is flat, i.e., $({\bar{R}^\alpha}_{~\beta\mu\nu}  = 0)$, which cancels spurious spatial terms that appears in the torsion scalar when the spin connection is set to vanish. Now, finding the connection, the torsion tensor together with the contortion tensor and the super-potential in view of equations \eqref{Gm}, \eqref{2.6} and \eqref{2.7} respectively, the torsion scalar may be computed using \eqref{2.8} to yield $\mathrm{T} = -6H^2$ for the spherical tetrad as well.\\

The quadratic torsion scalar so obtained is the same in either case and it replaces the Ricci scalar appearing in the `Einstein-Hilbert (EH) action. But as already mentioned, the field equations in that case are equivalent to GTR (TEGR) and no new physics emerges. Nonetheless, as in the modified theory of gravity, if an arbitrary functional form $f(\mathrm{T})$ is considered, the action without spin connection (Cartesian coordinates), takes the following form,
\begin{eqnarray}\label{2.9} A_\mathrm{T} = \int||e|| f(\mathrm T)d^4 x +\mathcal{S}_m,\end{eqnarray}
where $||e|| = |\mathrm{det} (e^a_\mu)| = \sqrt{-g}$ and $\mathcal{S}_m$ is the matter action. 
On the contrary, as the spin connection is associated, the action takes the form
\begin{eqnarray}\label{2.9b} A_{\mathrm{T}(e,\omega)} = \int||e|| f({\mathrm T}(e,\omega))d^4 x +\mathcal{S}_m,\end{eqnarray}
in which both the tetrads as well as the spin connection parametrized by the Lorentz metrics are considered, admits LLI \cite{Tcov1}. The above actions \eqref{2.9} as well as \eqref{2.9b} are the `generalized version of metric teleparallel gravity theory (GMTG)', which deviates from GTR to a great extent. The associated field equations are,
\begin{eqnarray}\label{2.10}\begin{split}& \left[e^{-1}\partial_\mu\left(e{e^\alpha}_a{S_\alpha}^{\nu\mu}\right) - {e^\mu}_a S^{\alpha\beta\nu}\mathrm{T}_{\alpha\beta\mu}\right]f_{,\mathrm{T}} + {e^\alpha}_a{S_\alpha}^{\nu\mu}\partial_\mu \mathrm{T} f_{,\mathrm{TT}} + {1\over 2}{e^\nu}_a f(\mathrm{T}) = {1\over 2}e^{-1} {T^\nu}_a,\end{split}\end{eqnarray}
where, $f_{,\mathrm{T}}$ and $f_{,\mathrm{TT}}$ are the first and the second derivatives of the arbitrary function $f(\mathrm{T})$ with respect to $\mathrm{T}$. The energy-momentum tensor in terms of the tetrad is expressed as, $T^{\mu\nu} = \eta^{ab} {T^\nu}_a {e^\mu}_b$, where ${T^\nu}_a = e^{-1} {\delta \mathcal{L}_m\over \delta {e^a}_\nu}$. In the flat Rw metric \eqref{RW} under consideration, since the quadratic scalar being identical, the field equations also remain unaltered, since the connection variation equation vanishes trivially in the second case (spherical coordinates). But, as mentioned, the former falls short of important physical requirement, violating LLI.\\ 

Lately, a manifestly LLI action was also presented associating a non-minimally coupled scalar field $(\phi)$, in the most general form of the action \cite{Tcov2},
\begin{eqnarray}\label{Tcov1} A_{\mathrm{T},\omega}[{e^a}_\mu, {\omega^a}_{b\mu}, \phi]= \int[f(\mathrm{T},\phi) - Z(\phi)\phi^{,\mu} \phi_{,\mu}]||e|| d^4x + \mathcal{S}_m,\end{eqnarray}
and is dubbed as `scalar-torsion theory'. In the above action ${\omega^a}_{b\mu}$ stands for non-trivial spin. In this case, the field equations found under the variation of the action with respect to the tetrads and the scalar field must satisfy the following connection-variation equation,
\begin{eqnarray} \label{con} \partial_\mu f_{,\mathrm T}\left[\partial_\nu\left(h {h_{[a}}^\mu {h_{b]}}^\nu + 2h h_c^{[\mu} {h_{[a}}^{\nu]} {\omega^c}_{b]}\right)\right] = 0\end{eqnarray}
Nonetheless, in the homogeneous and isotropic RW metric, the tetrad and the scalar field equations trivially satisfy the spin connection variation equation for spatial curvature $k = 0, \pm 1$ \cite{Tcov2}. Let us again mention that the spin connection does not give any independent dynamical equation, rather it acts as pure gauge maintaining LLI, accounting inertial effects.  

\subsection{Symmetric Teleparallel theory of gravity:}
Some serious issues in connection with LLI and strong coupling problems were encountered initially while handling generalized teleparallel gravity theory with torsion $f(\mathrm{T})$. Therefore, yet another formalism of teleparallel gravity theory emerged in the curvature-free $({\tilde{R}^\alpha}_{~\beta\mu\nu}  = 0)$ and torsion-free $({\mathrm{T}_{\mu\nu}}^\lambda = 0)$ environment. This is often referred to as the symmetric teleparallel theory of gravity. In symmetric teleparallel gravity, the vanishing curvature constraint imposes the connection to be purely inertial, i.e., it differs from the trivial connection by a general linear gauge transformation. Here, the quadratic non-metricity scalar $Q$ plays the role of the Ricci-scalar in the Einstein-Hilbert (EH) action, and differs from GTR by a total derivative term and therefore no new physics emerges. Recent focus is therefore on $f(Q)$ theory, which is also called, `the generalized symmetric teleparallel gravity theory' (GSTG). Here, we delineate the geometric structure of symmetric teleparallel gravity theory. Given a metric tensor  $g_{\mu\nu}$, the only non-trivial object associated to the connection is the non-metricity tensor,
\begin{eqnarray}\label{2.11} Q_{\alpha\mu\nu} = \nabla_\alpha g_{\mu\nu} = g_{\mu\nu,\alpha} - g_{\nu\rho}{\Gamma^\rho}_{\mu\alpha} -g_{\rho\mu}{\Gamma^\rho}_{\nu\alpha} \ne 0,\end{eqnarray}
where $\nabla$ is the covariant derivative satisfying curvature-free and torsion-free conditions. Since the covariant derivative of the metric tensor does not vanish, therefore the term non-metricity is used to identify the tensor and the quadratic scalars of symmetric teleparallel gravity. Two distinct traces of non-metricity vectors can be constructed, viz.,
\begin{eqnarray}\label{2.12} Q_\alpha= g^{\mu\nu}Q_{\alpha\mu\nu} = {Q_{\alpha\nu}}^\nu;\;\;\;\tilde{Q}_\alpha= g^{\mu\nu}Q_{\mu\alpha\nu} = {Q_{\nu\alpha}}^\nu,\end{eqnarray}
in view of which one can built a non-metricity conjugate, also called the superpotential tensor as,
\begin{eqnarray}\label{2.13} P^{\alpha\mu\nu} = -{1\over 4}Q^{\alpha\mu\nu} +{1\over 2}Q^{(\mu\nu)\alpha} +{1\over 4}(Q^\alpha + \tilde{Q}^\alpha)g^{\mu\nu} -{1\over 4}g^{\alpha(\mu}Q^{\nu)},\end{eqnarray}
where the last term is trace-dependent correction that ensures the super potential is properly conjugate to the non-metricity tensor. Finally, the quadratic non-metricity scalar $Q$ is found as,
\begin{eqnarray} \label{2.14} Q = -\frac{1}{4}Q_{\alpha\mu\nu}Q^{\alpha\mu\nu} + \frac{1}{2}Q_{\alpha\mu\nu}Q^{\mu\nu\alpha} + \frac{1}{4}Q_\mu Q^\mu - \frac{1}{2}Q_\mu \hat{Q}^\mu = R + \nabla_{\alpha}(Q^{\alpha}-{\tilde{Q}}^\alpha),\end{eqnarray}
which is constructed from the curvature-free and torsion-free non-metricity tensor. As mentioned, under replacement of the curvature scalar by the non-metricity scalar in the EH action, simply STEGR emerges since the term associated with the Ricci scalar is the total derivative term which does not contribute in the dynamics and so no new physics is expected. Nonetheless, for an arbitrary functional form $f(Q)$ the action
\begin{eqnarray} \label{2.15} A[Q,\Gamma] = \int f(Q)\sqrt{-g} d^4 x +\mathcal{S}_m,\end{eqnarray}
represents the `generalized version of symmetric teleparallel  theory of gravity' (GSTG), which again deviates from GTR to a great extent. The above action \eqref{2.15} when varied with respect to the metric tensor $g^{\mu\nu}$ and the connection ${\Gamma^\alpha}_{\mu\nu}$, following field and connection variation equations respectively emerge,
\begin{eqnarray} \label{2.16} \begin{split}&
\frac{2}{\sqrt{-g}} \nabla_\lambda (\sqrt{-g}f_Q{P^\lambda}_{\mu\nu}) +\frac{1}{2}f g_{\mu\nu} + f_Q(P_{\nu\rho\sigma} Q_\mu{}^{\rho\sigma}
-2P_{\rho\sigma\mu}{Q^{\rho\sigma}}_\nu) = -\kappa T_{\mu\nu},\\&
\nabla_\mu\nabla_\nu\left(\sqrt{-g} f_{,Q} {P^{\mu\nu}}_\lambda\right) = 0.\end{split}\end{eqnarray}
The Lie derivative of the connection with respect to the isometry vector ($Y$) associated with a space-time metric vanishes, i.e.,
\begin{eqnarray}\label{2.17} \mathcal{L}_Y{\Gamma^\mu}_{\alpha\beta} =& Y^\rho {\partial{\Gamma^\mu}_{\alpha\beta}\over \partial x^\rho} + {\Gamma^\mu}_{\rho\beta}{\partial Y^\rho\over \partial x^\alpha} + {\Gamma^\mu}_{\alpha\rho}{\partial Y^\rho\over \partial x^\beta} - {\Gamma^\rho}_{\alpha\beta}{\partial Y^\mu\over \partial x^\rho} + \frac{\partial^2 Y^\mu}{\partial x^\alpha\partial x^\beta} = 0.\end{eqnarray}
This equation, in the curvature-free and torsion-free environment, is used to find all possible connections involved in a space-time metric. Nonetheless, for curvature-less and torsion-less conditions, the affine connection may also be expressed as
\begin{eqnarray}\label{2.18} {\Gamma^\alpha}_{\mu\nu} = {\partial {x^\alpha}\over \partial\xi^\rho}\left({\partial^2 \xi^\rho\over \partial x^\mu\partial x^\nu}\right),\end{eqnarray}
where, the four scalar fields $\xi^\rho (x^\alpha)$ are arbitrary function of $x^\alpha$ and are called the St\"ueckelberg fields associated with the diffeomorphism. Under a special choice $\xi^\rho (x) = x^\rho$, the above connection being the derivative of the delta function, vanishes. This choice is called the coincidence gauge for which covariant derivative leads to partial derivative\footnote{In the torsion-less and curvature-less environment, the affine connection is symmetric in lower indices and flat, which can be transformed away locally. So there always exists a coordinate $\xi^\rho$, such that ${\Gamma^\rho}_{\mu\nu}(\xi)=0$ which is called the coincidence gauge. Now under a coordinate transformation $\xi^\rho = \xi^\rho(x)$, the transformed affine connection reads as, ${\Gamma^\alpha}_{\mu\nu}(x) = {\partial x^\alpha \over \partial \xi^\rho}{\partial \xi^\sigma \over \partial x^\mu}{\partial \xi^\lambda\over \partial x^\nu}{{\Gamma^\rho}_{\sigma\lambda}(\xi)} + {\partial {x^\alpha}\over \partial\xi^\rho}\left({\partial^2 \xi^\rho\over \partial x^\mu\partial x^\nu}\right)$. Since the first term vanishes, so \eqref{2.18} emerges. When $\Gamma$ is treated as independent geometric object, under LLT, non-metricity tensor transforms covariantly, while the scalar $Q$ remains invariant and hence the action. Covariant symmetric teleparallel gravity separates inertial effects from veritable gravitational degrees of freedom. Choosing coincidence gauge sets $\Gamma =0$ and as a result, compensating inertial connection is withdrawn.}, i.e.,
\begin{eqnarray}\label{2.19} Q_{\alpha\mu\nu} = \nabla_\alpha g_{\mu\nu} = g_{\mu\nu,\alpha},\end{eqnarray}
and the metric $(g_{\mu\nu})$ becomes the only independent variable, which had been widely used in the past for its simplicity. Nonetheless, in coincidence gauge one sacrifices LLI. In the covariant formulation of the theory, affine connection is treated as an independent geometric object which then becomes a pure gauge connection \eqref{2.18}. As both the Riemann curvature tensor and torsion tensor vanish, so the connection contains only inertial information. Treating the affine connection as independent variable the covariant formulation of GSTG and such a LLI action reads as \cite{Qcov1, Qcov2},
\begin{eqnarray} A_{Q,\Gamma}= \int f(Q)\sqrt{-g} d^4 x +\mathcal{S}_m.\end{eqnarray}

Similar to the scalar-tensor theory of GTR, it is also possible to construct a scalar-non-metricity theory in the following form
\begin{eqnarray}\label{Action} A_{Q,\Phi} = \int [f'(\Phi) Q + f(\Phi) - \Phi f'(\Phi)]\sqrt{-g} d^4x + \mathcal{S}_m, \end{eqnarray}
where, $\Phi$ is an auxiliary variable and prime represents the derivative with respect to $\Phi$. However the above action again falls short of LLI and so the covariant formulation of the theory is formulated through the following scalar-vector-tensor formalism \cite{Qcov1}, 
\begin{eqnarray} \label{Qcov1}\begin{split}& A_{\mathrm{Cov}} = \int \left[{1\over 2} \left\{f(\Phi) Q + (Q^\alpha - \hat{Q}^\alpha)\partial_\alpha f(\Phi)\right\} - \omega(\Phi)\partial_\alpha \Phi \partial^\alpha \Phi - U(\Phi)\right] \sqrt{-g} d^4 x + \mathcal{S}_m,\\&
\mathrm{where,}~~Q^\alpha = {Q^{\alpha\mu}}_\mu\;\;\;\&\;\;\;\hat{Q}^\alpha = {Q_\mu}^{\mu\alpha}.\end{split}\end{eqnarray}
The very presence of the term $(Q^\alpha - \hat{Q}^\alpha)\partial_\alpha f(\Phi)$ makes the action \eqref{Qcov1} manifestly covariant for non-zero affine connections. In fact, in the spatially flat RW metric \eqref{RW}, it simply gives non-minimally coupled scalar-tensor theory of gravity for the first connection, which is manifestly covariant, as we shall exhibit in section 6, while for other three connection, it is modified and different cosmological evolution is expected.\\

To this end let us mention that the connection variation equation gives constraints. In the RW space-time, apart from the coincidence gauge, in which all the connections vanish trivially, four different connection variation equations (three for the spatially flat and one for non-flat space-times) are generated in covariant formalism.

\section{Looking through the vacuum dominated era:}
As mentioned, we aim at exhibiting some additional features of these theories. Firstly, let us mention that we are typically working in flat FLRW background choosing cosmic time gauge, setting the lapse function $\mathrm{N} = 1$ \eqref{RW}, since all the connections admit diffeomorphic invariance \cite{DA, DA2}. Now, as being artefact of gauge theory, teleparallel theories admit different connections for a particular space-time metric corresponding to different gauge choice. In the spatially flat RW metric under consideration \eqref{RW}, altogether there exists four different sets of affine connections for symmetric teleparallel theory under different gauge choice. In coincidence gauge, affine connections vanish yielding $Q = -6H^2$, while for the following non-trivial affine connections,

\begin{eqnarray}\label{Hoh11} \begin{split}& {\Gamma^0}_{00} = \gamma(t),\;\;{\Gamma^1}_{22} = -r,\;{\Gamma^1}_{33} = - r\sin^2{\theta};\\& {\Gamma^2}_{12} = {\Gamma^2}_{21} ={1\over r},\;\;\; {\Gamma^2}_{33} = -\sin{\theta}\cos{\theta};\\&
{\Gamma^3}_{13}={\Gamma^3}_{31} = {1\over r},\;{\Gamma^3}_{23} = {\Gamma^3}_{32} = \cot{\theta};\;\;.\end{split}\end{eqnarray}
one again finds $Q = -6H^2$ and the arbitrary parameter ($\gamma$) remains inert. Therefore both give identical field equations. The connection variation equations for the other two connections in spatially flat space ($k = 0$) simply yield $f_{,QQ} = 0$ and therefore cannot be generalized, resulting in STEGR \cite{DA2}. In this sense, $Q = - 6H^2$ is the only possibility for GSTG in the spatially flat ($k = 0$) case. Further, since in spatially flat $k = 0$ case a LLI RW space-time also reads as $\mathrm{T} = -6H^2$, and the field equations remain unaltered from the two viable cases for symmetric teleparallel gravity theory \footnote{As mentioned, in the case of flat FLRW metric finding the correct spin connection is not an issue and the torsion scalar as well as dynamics remains unaltered.}, therefore we can consider $\mathrm{T} = Q = X = -6H^2$, on the same footing, without any loss of generality. We therefore start with the following generalized teleparallel gravitational action,
\begin{eqnarray}\label{A} A = \int\left[f(X)+L_m \right]{\sqrt{-g}} d^4 x,\end{eqnarray} 					
and as mentioned, $X = - 6H^2 = - 6{{\dot a}^2\over a^2}$ stands for either $\mathrm{T}$ or $Q$, while $L_m$ stands for the matter Lagrangian. Treating $X +{6{\dot a^2}\over a^2} =0$, as a constraint and introducing it through a Lagrange multiplier $\lambda$, action \eqref{A} is expressed as,

\begin{eqnarray}\label{A1} A=\int\left[f(X)- \lambda\left(X+ {6{\dot a}^2\over a^2}\right) + L_m\right]{a^{3}} dt,\end{eqnarray}
absorbing $2 \pi^2$, arising due to the integration over the three-space in the action. Under the variation of the action with respect to $X$, $\lambda = f'(X)$ emerges, where $f'(X)$ is the derivative of $f(X)$ with respect to $X$. Substituting it back in \eqref{A1}, the action finally reads as,

\begin{eqnarray}\label{A3} A=\int\left[\Big\{f(X)- X f'(X)\Big\}{a^{3}}-{6a{\dot a}^2}f'(X) + L_m a^{3}\right] dt,\end{eqnarray}
and correspondingly the point Lagrangian can be expressed as,

\begin{eqnarray}\label{L} L = \Big[f(X)- X f'(X)\Big]{a^{3}}-{6a{\dot a}^2 }f'(X) + L_m a^{3}.\end{eqnarray} 	
The field equations are,

\begin{eqnarray}\label{F1} 12 H^2f'(X) + f(X)= \sum\rho_i,\end{eqnarray}								
\begin{eqnarray}\label{F2} 4{\dot H}\left[12H^2 f''(X) - f'(X)\right] - 12H^2f'(X) - f(X) = \sum p_i,\end{eqnarray}
where $\sum \rho_i$ and $\sum p_i$ are the energy density and the pressure of some fluid components respectively. In the very early pure vacuum dominated era, $(\sum\rho_i = \sum p_i = 0)$ the field equations \eqref{F1} lead to,

\begin{eqnarray} f(X) = f_0 \sqrt{X} = f_0\sqrt{-6 H^2}.\end{eqnarray}
Hence, either the Hubble parameter $(H)$ or the function $f(X)$ must be imaginary. Therefore the suggestion that $f(Q) = Q - 6\lambda H^2\left(Q\over 6 M^2\right)^\delta$ can probe early inflation, when $H >> M$ and $\delta > 0$, $M$ being an arbitrary scale \cite{coincident} is definitely not true. So unlike curvature induced inflation in $f(R)$ theory \cite{Staro}, torsion or non-metricity (inclusive of higher degree terms) induced inflation in teleparallel gravity remains obscure in general for $f(Q)$ and $f(\mathrm{T})$ gravity in RW space-time. Clearly, inflation in the very early universe in teleparallel gravity may only be driven by a scalar field. Of-course, if a viable inflation and graceful exit from inflation are realizable, then hot big-bang is procured following particle production either due to oscillation of the scalar field or under instantaneous reheating (depending on the potential), whence radiation dominated era initiates. With this precursor, we now first choose a suitable form of $f(X)$ and recapitulate the issue of `Branched Hamiltonian' studied earlier in \cite{IJJ}.

\section{The issue of Branched Hamiltonian:}

Since teleparallel theories do not admit arbitrary forms in the vacuum dominated era, so to study the astrophysical aspects and cosmological evolution of the universe, a reasonable form of $f(X)$ is required. Here we take $f(X)=\alpha_1 X + \beta_1 X^2$, as minimal extension over GTR, where $\alpha_1$ and $\beta_1$ are constant coupling parameters. Note that in the late stage of cosmic evolution, the second term becomes negligible and under the choice $\alpha_1 = {1\over 16\pi G} = 0.5 M_P^2$, GTR is recovered which clearly passes solar test. Further, such a form being an outcome of reconstruction programme, has been extensively used in the literature particularly in the very early universe to explore its effect on the scalar field driven inflationary dynamics and in some cases to study late stage of cosmic evolution \cite{In1, In2, In3, In4, In5, In6, ln7}. It has also been revealed that the inflationary parameters educed from such a form when associated with a scalar field, are in perfect agreement with the recently published Plancks inflationary data \cite{ln7, MNSA}. Nonetheless, late-time cosmic evolution requires yet another term, such as $-6\lambda M^2\left({X\over 6M^2}\right)^\delta$, with $\delta < 0$, while $\lambda$ is a dimensionless parameter and $M$ is some scale required to unify early inflation $H \gg M$ with late-time accelerating phase $H \ll M$ \cite{coincident}. As we have already noticed that higher-degree terms cannot be associated in the pure vacuum era in the absence of a scalar field, so the scale may be removed. A suggestive complete form is therefore $f(X) = \alpha_1 X + \beta_1 X^2 + \eta{X}^\delta$, ($\delta < 0$), which might be suitable to mimic the dark energy issue. Nevertheless, in the very early universe, which is our present concern, the last term remains sub-dominant and so we neglect this term for simplicity. In the following, we particularly aim at the deliberation of the issue of `Branched Hamiltonian' revealed earlier \cite{IJJ} with such a form of $f(X) = \alpha_1 X + \beta_1 X^2$ and leaving out the scalar field, since it is irrelevant to the analysis. 

\subsection{Analyzing previous result on branching:}
After the transition from the quantum domain $(l_P = 10^{-43}~s)$, the universe is supposed to appear as `vacuum dominated era $(\rho = p = 0)$', where, $\rho$ and $p$ are the energy-density and the thermodynamic pressure for the perfect fluid. Our previous analysis suggest that a scalar field must be associated in the early universe, but we temporarily suppress it, since it plays no role in the issue of branching. The point Lagrangian for the above form of $f(X)$ reads as,

\begin{eqnarray} L=-{6\alpha_1{a\dot a}^2}+\frac{36\beta_1{\dot a}^4}{a}.\end{eqnarray}
The canonically conjugate momentum and the Hamiltonian with respect to scale factor `$a(t)$' are,

\begin{eqnarray} p_{a}=-{12\alpha_1{a\dot a}}+\frac{144\beta_1{\dot a}^3}{a},~~~~\mathrm{H}= E =-{6\alpha_1{a\dot a}^2}+\frac{108\beta_1{\dot a}^4}{a}. \end{eqnarray}
Note that, since the Hamiltonian $(\mathrm{H})$ cannot be cast in terms of the phase-space variables, so we have also used $E$ for the energy notation. Now, to scrutinize the system, we can choose any instant $0 < a(t) \le 1$,  after the present instant is set as, $a(t_0)=1 $, so that

\begin{eqnarray}\label{Br}p_{a}=-{12\alpha_1{\dot a}}+{144\beta_1{\dot a}^3},~~~~\mathrm{H}=E=-{6\alpha_1{\dot a}^2}+{108\beta_1{\dot a}^4}.\end{eqnarray}
Under the choice, $\alpha_1 > 0,~~\beta_1 > 0$, $p_{a}$ versus $\dot a(t)$ and $\mathrm{H}/E$ versus $p_{a}$ plots are represented graphically in fig-1 and fig-2 (with $\alpha_1 = {M_P^2\over 2} = 0.5$, in the unit $8\pi G = M_P^{-2} = 1$ and $144\beta_1 = 2.5$) respectively, which were also demonstrated earlier in \cite{IJJ} with different values of $\alpha_1$ and $\beta_1$.\\

\begin{figure}
\begin{minipage}[h]{0.45\textwidth}
\centering
\includegraphics[ width=1.10\textwidth] {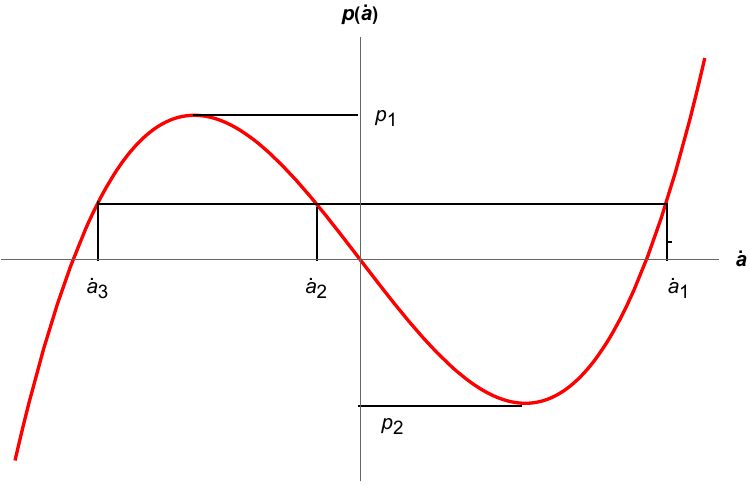}
 \caption{The figure depicts multivaluedness of $\dot a$, for $\alpha_1 > 0,~\beta_1 > 0$.}
      \label{fig:1}
   \end{minipage}%
  \hfill
\begin{minipage}[h]{0.40\textwidth}
\centering
\includegraphics[ width= 1.10\textwidth] {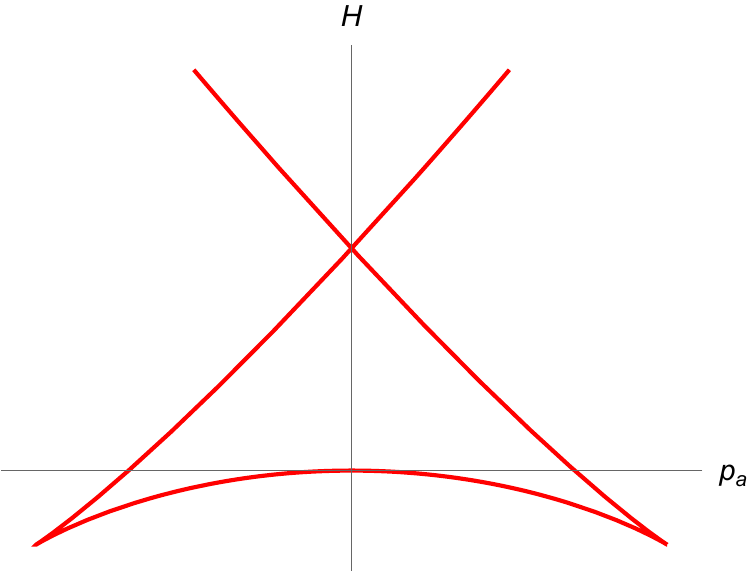}
 \caption{Branched Hamiltonian with cusp is depicted.}
      \label{fig:2}
   \end{minipage}%
\end{figure}

Fig-1 depicts that in the interval $[p_1, p_2]$, the velocity $(\dot a )$ is multi-valued and therefore it is not possible to predict the initial value of the velocity $(\dot a)$ the momentum belongs to. As a result, the equations of motion switch or jump instantaneously from one value of the velocity $(\dot a)$ to another, because these sudden jumps leave $p_a$ unaltered and also comply with the field equations. Additionally, fig-2 suggests that $\mathrm{H}$ or alternatively $E$ is also multi-valued and at any instant of time it is not possible to decide which `branch' of the Hamiltonian/Energy one should consider. Consequently, the associated field might propagate for a while with one choice of the Hamiltonian and then switch over to another and so on. Since such switching occurs within arbitrarily small interval of time, the classical motion is visualized as a series of erratic wiggles, which happens in an unanticipated manner. Hence, the behaviour of the system described by the action \eqref{A} remains uncertain for a range of initial non-vanishing data. Such a situation arises because the Lagrangian is quartic in the velocities and therefore expressions for velocities are multi-valued functions of momentum, resulting in the so-called `multiply branched Hamiltonian with cusps' (fig-2). Due to such jump from one branch of the Hamiltonian to the other at any instant of time, the classical solution remains unpredictable. Note that the existence of cusps restricts the domain of the variables.\\

The pathology of Branched Hamiltonian appears in different fields of physics and attempts to resolve the issue remains unsuccessful over decades. In the context of a toy model of Lanczos-Lovelock gravity in particular, it had been shown that different Hamiltonians emerge from different attempts made to circumvent the problem \cite{15}. Nonetheless, in the absence of a feasible `unique theory', it had been demonstrated that the pathology may be bypassed by adding a
$\ddot q^2$ term in the Pais-Uhlenbeck oscillator action \cite{16} and $R^2$ term in the Lanczos-Lovelock action \cite{15,16}. The same technique was adopted in $f(\mathrm{T})$ theory of gravity earlier \cite{IJJ}. The issue of branching is true for any GMTG and GSTG, once the form of $f(X) = \alpha_1 X + \beta_1 X^2$ is chosen, while additional terms do not affect the issue. However, it is worth mentioning that the issue of branching significantly depends on the choice of the signature of the parameters $\alpha_1$ and $\beta_1$, which are taken positive arbitrarily \cite{IJJ}. While, the signature is fixed once and forever from the very beginning in the Pais-Uhlenbeck and the Lanczos-Lovelock actions, it is not so in the present case. One should rather fix the issue in view of the the energy condition based on earlier works \cite{MNSA}, which was not considered previously \cite{IJJ}. In the following, we consider the energy condition extensively to fix the signatures of $\alpha_1$ and $\beta_1$.

\subsection{Energy conditions, the need for a scalar in the very early universe:}

The energy conditions for $f(Q) =\alpha_1 Q + \beta_1 Q^2$ has been studied recently in \cite{ln7}, where it has been revealed that one needs to fix $\alpha_1 > 0$ and $\beta_1 < 0$, if a minimally coupled scalar field is associated. Let us recall that the modified and alternative (teleparallel) theories of gravity are contemplated as alternatives to the uncanny dark energy and the motivation of these theories is to address the cosmic riddle in the late universe without dark energy. Now, after graceful exit from inflation, particles are produced commencing a hot big-bang era. This is the epoch dominated by radiation. It may be possible that a tiny amount of scalar field remains in the radiation era too, which is redshifted away leaving no trace in the matter dominated era. However, a much stringent situation emerges, if the scalar field is completely wiped out in the process of producing particles, so that the radiation era consists of perfect fluid only (inclusive of dark matter). Later with the expansion, the universe appears to be like pressure-less dust era. In these situations, the energy density $\rho$ and the thermodynamic pressure $p$ are related by $p = \omega  \rho$, in the present choice of unit $c =1$. In essence, it is necessary to investigate the energy condition in the presence of barotropic fluid for which $p \ge 0, \rho > 0$ have to be satisfied. These in turn automatically satisfy all the four energy conditions. Now, under the choice $f(X) =\alpha_1 X + \beta_1 X^2$,  equations \eqref{F1} and \eqref{F2} read as,
\begin{eqnarray} 6\alpha_1 H^2-108\beta_1 H^4=\rho ~~~\text{and}~~~ -6\alpha_1 H^2-4\alpha_1{\dot H} +144\beta_1{\dot H}H^2+108\beta_1 H^4 = p .\end{eqnarray}
Clearly $\alpha_1 > 0$ and $\beta_1 < 0$, ensures $\rho > 0$. We therefore replace $\alpha_1$ by $\alpha$ and $\beta_1$ by $-\beta$, where both $\alpha > 0$ and $\beta > 0$. Let us now inspect the pressure equation, which now reads as,
\begin{eqnarray} -6\alpha H^2-4\alpha{ \dot H} - 144\beta{\dot H}H^2 - 108\beta H^4 = p.\end{eqnarray}
Now, since in the expanding model, $\dot H \le 0$, so $p > 0$ implies, $4(\alpha + 36 \beta H^2)|\dot H| > 6H^2(\alpha + 18 \beta H^2)$, requiring $|\dot H| < {3\over 2} H^2$. Therefore for the choice $a = a_0 t^n$, the condition finally requires $n > {2\over 3}$, and a decelerated expansion is realizable, which is the viable cosmic evolution. Thus, the energy conditions fix the coupling constants $\alpha$ and $\beta$ and the form we shall deal with is,
\begin{eqnarray} \label{finalf}f(X) = \alpha X - \beta X^2,\end{eqnarray}
where, $\alpha > 0$ and $\beta > 0$. Now let us consider even early universe where inflation is driven by a scalar field having components $\rho_\phi = {1\over 2}\dot \phi^2 + V(\phi)$ and $p_\phi = {1\over 2}\dot \phi^2 - V(\phi)$. In that case, `the weak energy condition' $\rho_\phi + p_\phi = - 4\alpha \dot H - 144 H^2 \dot H > 0$ is satisfied, since in the expanding model $\dot H < 0$. Nonetheless, the strong energy condition reads as,
\begin{eqnarray} \rho_{\phi} + 3p_{\phi} = 12[\alpha|\dot H| + 36 \beta H^2 |\dot H| - \alpha H^2 - 18 \beta H^4].\end{eqnarray}
Clearly for a quasi de-Sitter solution $a = a_0 e^{H(t)t}$, where the `Hubble parameter $H$' is a slowly varying function of $t$, the last term dominates, and strong energy condition appears to be violated. But that is not an issue, since it is possible for $V(\phi) > \dot \phi^2$, which is the limiting slow-roll condition.\\

In view of the energy conditions we now consider the form \eqref{finalf}, in which both the parametric signatures of $\alpha > 0, \; \beta > 0$ are now fixed, and muse on to the issue of Branched Hamiltonian yet again. The Lagrangian, the momentum and the energy now read as,

\begin{eqnarray}\label{LPH} L=-{6\alpha{a\dot a}^2}-\frac{36\beta{\dot a}^4}{a},~~~~
p_{a}=-{12\alpha{a\dot a}}-\frac{144\beta{\dot a}^3}{a},~~~~~~\mathrm{~H}= E =-{6\alpha{\dot a}^2}-{108\beta{{\dot a}^4\over a}}, \end{eqnarray}
and as before, for the choice for $a(t_0) = 1$, we present graphical representation of the momentum versus velocity and the Hamiltonian/Energy versus momentum as depicted in fig-3 and fig-4. Clearly, the velocity as well as the Hamiltonian are now single valued and the pathology of Branched Hamiltonian disappears in view of the energy conditions. This is the virtue of the second order teleparallel theory over Lanczos-Lovelock gravity. Next, considering the above form of $f(X) = \alpha X-\beta X^2$, with proper sign of $\alpha > 0, ~\beta > 0$, the $(^0_0)$ equation and the `$a$' variation equation are expressed in the presence of a scalar field as,
\begin{eqnarray}\label{S1} 6\alpha H^2 + 108\beta H^4=\rho_\phi, ~~~6\alpha H^2 + 4\alpha \dot {H}+ 144\beta\dot{H}H^2 + 108\beta H^4=-p_\phi.\end{eqnarray}
\begin{figure}
\begin{minipage}[h]{0.37\textwidth}
\centering
\includegraphics[ width=1.10\textwidth] {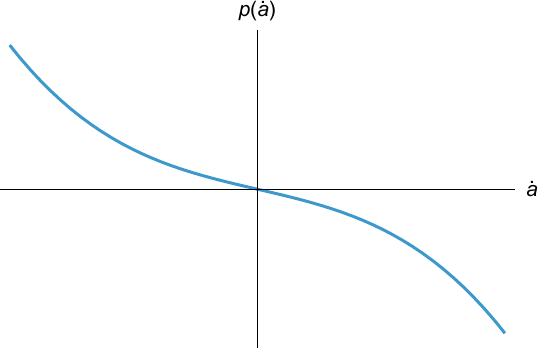}
 \caption{$\alpha >0,~\beta <0$ shows single value of $\dot a$ .}
      \label{fig:3}
   \end{minipage}%
  \hfill
\begin{minipage}[h]{0.37\textwidth}
\centering
\includegraphics[ width=1.10\textwidth] {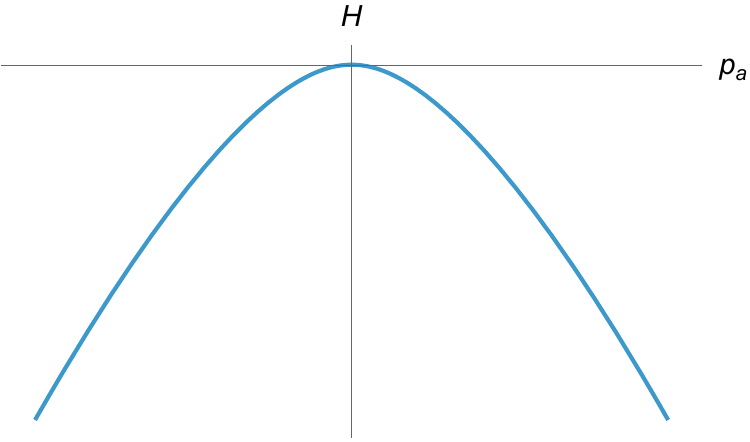}
 \caption{Single valued Hamiltonian.}
      \label{fig:4}
   \end{minipage}%
 \end{figure}
Let us clearly mention that Lanczos-Lovelock (LL) theory in 4-dimension (say), is based on a particular form of higher order gravity theory. Therefore, LL theory always suffers from the issue of Branched Hamiltonian. In contrast, $f(T/Q)$ gravity theories may be associated with arbitrary forms and not all are inflicted with the issue, apart from the form like $f(X) = \alpha X + \beta X^n, n\ge 2$. If extended classes of teleparallel theories such as ($f(X,B)$) theory where $B$ is the boundary term, or $f(X, X_\mathcal{G})$ theory where $X_\mathcal{G}$ is the teleparallel Gauss-Bonnet term, also contain such term, the problem cannot be avoided. Nonetheless, we have shown that, the coupling parameter $\beta$ has to be negative ($\beta < 0$) to satisfy of energy condition and so the problem doesn't arise at all.

\subsection{\bf{Slow-roll inflation:}}
Before defying to work any further, the Planck satellite observations released a host of data-sets imposing tighter constraint on the inflationary parameters, viz. on the spectral index $(0.9631 \le n_s \le 0.9705)$ as well as on the tensor to scalar ratio $r < 0.055$ \cite{Planck1, Planck2}. Of late, combination of Planck PR4 data with ground-based experiments such as, BICEP/Keck 2018 (BK18), BAO and CMB lensing data, tightens the scalar to tensor ratio even further to $r < 0.032$ \cite{Planck3}. More recently in 2025, ground based `ACT' (Atacama Cosmology Telescope stationed at Atacama desert, Chile) released latest dataset. Combination of ACT DR6 (Data Release 6) data with Planck and BICEP/Keck (BK18) polarization, put constraints on $n_s \approx 0.9743\pm 0.0034$ and $r < 0.038$ \cite{ACT1, ACT2, ACT3}. Further, both Planck and ACT favoured single field slow-roll inflation for flat plateau-like potential. Earlier as mentioned, scalar field driven inflation had been studied in the context of gravity with torsion, for the choice $f(\mathrm{T}) = \alpha \mathrm{T} + \beta {\mathrm{T}}^2$, in the background of RW metric \cite{MNSA} and also for gravity with non-metricity in the background of anisotropic Bianchi-1 space-time for $f(Q) = \alpha Q - \beta Q^2$ \cite{ln7}, $\alpha > 0,~\beta > 0$ and excellent agreement were found. Here, we consider $f(X) = \alpha X - \beta X^2$ ($\alpha > 0, \beta > 0$) in the background of RW space-time for four different scalar field potentials, i) $V(\phi)=V_0+V_1\phi^2$,~ii) $V(\phi)=V_0+V_1\phi^4$,~ iii) $V(\phi)= V_0-{V_1\over \phi}$ and iv) $V(\phi)=V_0-{V_1\exp(-b\phi)}$, out of which the third and fourth represent flat potentials on the onset of inflation, when $\phi$ is large enough. The field equations \eqref{S1} now read as,
\begin{eqnarray}\label{S2} 6\alpha H^2+108\beta H^4={1\over 2}{\dot \phi}^2+V(\phi),\end{eqnarray}
\begin{eqnarray} 6\alpha H^2+4\alpha \dot{H}+144\beta\dot{H}H^2+108\beta H^4=-{1\over 2}{\dot \phi}^2+V(\phi),\end{eqnarray}
\begin{eqnarray}\label{KG} \ddot{\phi} +3H\dot\phi+V'(\phi)=0 ,\end{eqnarray}
where the third equation is the $\phi$ variation (Klein-Gordon) equation. Applying the standard slow-roll conditions
$\dot\phi^2\ll V(\phi)$ and $|\ddot\phi|\ll 3{H}|\dot\phi|$, equations (\ref{S2}) and (\ref{KG}) read as,
\begin{eqnarray}\label{com} \gamma H^4+ 6\alpha H^2- V(\phi)=0, ~~~~~~~ 3H\dot\phi+V'(\phi)=0,\end{eqnarray}
where $108\beta=\gamma$. In view of \eqref{com}, $H^2$ can be solved and we promptly obtain (considering the positive sign),

\begin{align} H^2 = \frac{\left[-3\alpha+ \sqrt{9\alpha^2+\gamma V(\phi)}\right]}{\gamma}.\label{soln}\end{align}
Additionally, combining equations (\ref{com}) one can compute the slow-roll parameters as,

\begin{align}\label{SR2}\begin{split} \epsilon={1\over 16\pi G}\bigg({V'\over V} \bigg)^2 \equiv - {12\alpha\dot {H}\over {H}^2}\bigg[\frac{3\alpha +\gamma H^2}{\big(6\alpha +\gamma H^2\big)^2}\bigg];\hspace{0.3 in}\eta = 2 \alpha \left({V''(\phi)\over V(\phi)}\right),\end{split}\end {align}
where, we consider $\alpha = {1\over 16\pi G}$. The relation of $\epsilon$ depicts that the potential slow roll parameter equals to the Hubble slow roll parameter, when $\gamma = 0$. Also, since $\frac{H}{\dot\phi}=-\frac{3\left[-3\alpha+\sqrt{9\alpha^2+\gamma V(\phi)}\right]}{\gamma V'(\phi)}$ in view of \eqref{S2}, one can compute the number of e-folds as,

\begin{align}\label{Nphi2} {N}(\phi)\simeq \int_{t_i}^{t_f}{H}dt=\int_{\phi_i}^{\phi_f}{{H}\over {\dot\phi}}d\phi= \int_{\phi_f}^{\phi_i}\frac{3\left[-3\alpha+\sqrt{9\alpha^2+\gamma V(\phi)}\right]}{\gamma V'(\phi)}d\phi.\end{align}
Any simple inflationary model gives two important independent observable parameters to describe the primordial power spectrum. These are the tensor to scalar ratio ($r$), which determines the shape of the potential and the spectral index ($n_s$), which determines the departure from scale invariance. One can compute the two from consistency relations and for the single scalar field inflation in GTR. In this process, these two turn out to be,
\be \label{rn}r = {P_T\over P_S} =16\epsilon;\;\;\;\;\;n_s -1 = {d\ln P_s\over dN}  =1-6\epsilon + 2\eta,\ee
where $P_T$ and $P_S$ are the tensor fluctuation and the scalar fluctuation being computed under perturbation of the tensor and the scalar modes respectively. Note that, since we are considering theory different from GTR only by a quartic term ($H^4$); linear perturbation remains unaffected and so the above forms \eqref{rn} may safely be considered here too.\\

\noindent
\textbf{Case-1:}\\
First, let us take into account the standard quadratic form of the potential, viz.,
\begin{eqnarray} \label{1V} V(\phi)= V_0+{V_1\phi^2}.\end{eqnarray}
The expressions for $\epsilon, ~\eta$ and $N$ in view of equations \eqref{SR2} and \eqref{Nphi2} are found respectively as,
\begin{align}\label{para5}\begin {split} &\epsilon=\alpha \bigg({V'\over V}\bigg)=\frac {4\alpha \phi^2}{\left({V_0\over V_1}+\phi^2\right)^2};\hspace{0.5 in}\eta =\frac{4\alpha}{\big({V_0\over V_1}+\phi^2\big)};\\& N=\int_{\phi_f}^{\phi_i}\frac{3\left[-3\alpha +\sqrt{9\alpha^2+\gamma V_1\bigg({V_0\over V_1}+\phi^2\bigg)}\right]}{2\gamma V_1\phi}d\phi.\end{split}\end{align}
In Table-1 and Table-2, we present two data sets for the  expressions \eqref{para5}, varying $\phi_i$ between $17 M_P \leq \phi_i \leq  20M_P$ and  ${V_0\over V_1}$ between $-30 M_P^2\leq {V_0\over V_1} \leq -1M_P^2 $, both of which show appreciable deviations from the experimental limit.
\begin{table}
\begin{minipage}[h]{0.48\textwidth}
      \centering
      \begin{tabular}{|c|c|c|c|}
     \hline\hline
      $\phi_i$ in $M_P$ & $n_s$ & $r$ & $ {N}$\\
      \hline
        17 &  0.9698 & 0.1188 & 46\\
        18 &  0.9733 & 0.1052 & 51\\
        19 &  0.9762 & 0.0938 & 56\\
        20 &  0.9787 & 0.0842 & 62\\
        \hline\hline
    \end{tabular}
     \caption{Inflationary parameters \eqref{para5} computed\\ for quadratic potential varying $\phi_i$ while fixing\\ ${\gamma V_1}= 0.03 M_P^2$, ${V_0 \over V_1}=-10 M_P^2$, $\alpha=0.5 M_P^2$ and \\$\phi_f= 3.9475M_p$. The observational limit set up \\for $r$ by Planck data is surpassed .}
      \label{tab:Table1}
   \end{minipage}
 \begin{minipage}[h]{0.48\textwidth}
      \centering
      \hspace{10.0 mm} \begin{tabular}{|c|c|c|c|c|}
     \hline\hline
      $V_0 \over V_1$& $\phi_f$ in $M_P$ in $M_P^2$ & $n_s$ & $r$ & ${N}$\\
      \hline
       -30 & 6.2297 & 0.9686 & 0.1199 & 48\\
       -20 & 5.2348 & 0.9711 & 0.1122 & 54\\
       -10 & 3.9475 & 0.9733 & 0.1052 & 60\\
       -1 &  1.9319 &  0.9751 & 0.0994 & 69\\
       \hline\hline
    \end{tabular}
     \caption{Inflationary parameters \eqref{para5} computed for quadratic potential varying $V_0 \over V_1$ (for which $\phi_f$ also varies automatically)  and fixing ${\gamma V_1}= 0.01 M_P^2$, $\alpha=0.5 M_P^2, ~\phi_i= 18M_p$. Here again data for $r$ shows substantial deviation set up by Planck.}
     \label{tab:Table2}
   \end{minipage}
   \end{table}
For the sake of completeness, we compute the energy scale of inflation in view of the relation \eqref{soln}, selecting one of the data presented in Table 1, viz., ($N = 46$, for which $\phi_i = 17M_P,~ \alpha= {1\over 16\pi G} = 0.5M_P^2, ~{{V_0\over V_1}=-10 M_P^2},~\gamma V_1= 0.03M_P^2$). Accordingly, we find,
\begin{align}\label{QdES}{H_*}^2 = {1.7588\over \gamma},\end{align}
which remains arbitrary unless $\gamma$ is fixed. The energy scale of inflation in a single scalar field model corresponding to GTR \cite{Wands}, is presented by the following expression,
\begin{align}\label{QQdES}{H_*} = 8\times 10^{13}\sqrt{r\over 0.2}GeV=6.166\times 10^{13}GeV =2.428\times 10^{-5} M_p,\end{align}
whose numerical value is calculated taking into account the value of the tensor-to-scalar ratio $r = 0.1188$ from the same data set ($N = 46$) of Table-1 under consideration. Thus, for $\gamma \approx 2.98 \times 10^{9}$ the scale of inflation \eqref{QdES} is comparable with the single field scale of inflation \eqref{QQdES}. Since $\gamma$ is fixed from physical consideration (sub-Planckian scale of inflation), the values of $\beta$, $V_1$ and $V_0$ may also be computed, which are,
\begin{align} \label{QdV2}\beta = 2.759\times 10^7,\hspace{0.2 in} V_1= 1.01\times 10^{-11}M_p^2, \hspace{0.2 in} V_0 = -10 V_1\approx -10.1\times 10^{-11}M_P^4.\end{align}
Finally, to exhibit graceful exit from inflation, we recall equation \eqref{S2}, which considering the above form of the potential, $V(\phi)=V_0+{V_1\phi^2}$, is written as,
\begin{align}\label{QdH3}-\frac{\gamma H^4}{V_1} - \frac{6\alpha H^2}{V_1}+\left[{\dot\phi^2\over 2V_1}+\left({V_0\over V_1}+{\phi^2}\right)\right]=0.\end{align}
During inflation, $H$ is slowly varying and all the terms in \eqref{QdH3} almost are of the same order of magnitude. However, as inflation ends the Hubble parameter decreases sharply, and so both the left hand side terms ${\gamma H^4\over V_1}$ and ${\alpha H^2\over V_1}$ may be neglected without sacrificing generality to find,
\begin{align}{\dot{\phi}}^2=-2\left[V_0+V_1\phi^2\right].\label{osc3}\end{align}
Taking into account, $\phi_f=3.9475M_P$, $V_0= -1.01\times 10^{-10}M_P^4$ and $V_1= 1.01\times 10^{-11}M_P^2$ from Table-1, it is possible to demonstrate that the above equation portrays oscillatory behaviour
\begin{align}\phi=\exp({i\omega t}),\end{align}
provided, $\omega \approx 2.689 \times 10^{-6}M_P$. Although, all the issues look good, but the inflationary parameters, particularly the tensor to scalar ratio $(r)$ does not quite agree with observation. In fact, to bring it down, the spectral index $(n_s)$ increases and goes far beyond observational limit. Therefore, quadratic potential is ruled out and it may be mentioned that Planck's observation ruled out quadratic potential too.\\

\noindent
\textbf{Case-2:}\\
Next, we consider the quartic potential,
\begin{eqnarray} \label{V} V(\phi)= V_0+{V_1\phi^4},\end{eqnarray}
and compute the inflationary parameters. The expressions for $\epsilon, ~\eta$ and $N$ in view of equations \eqref{SR2} and \eqref{Nphi2} are found respectively as,
\begin {align}\label{para4}\begin {split} &\epsilon=\alpha \bigg({V'\over V}\bigg)=\frac {16\alpha \phi^6}{\left({V_0\over V_1}+\phi^4\right)^2};\hspace{0.5 in}\eta =\frac{24\alpha\phi^2}{{V_0\over V_1}+\phi^4};\\& N=\int_{\phi_f}^{\phi_i}\frac{3\left[-3\alpha +\sqrt{9\alpha^2+\gamma V_1\bigg({V_0\over V_1}+\phi^4\bigg)}\right]}{4\gamma V_1\phi^3}d\phi.\end{split}\end{align}
In Table-3, we display a data set for the  expressions \eqref{para4}, varying $\phi_i$ between $28 M_P \leq \phi_i \leq  38M_P$, which again shows substantial deviation from the experimental limit particularly in the value of the tensor to scalar ratio. In fact, any attempt to reduce it, increases the spectral spectral index beyond experimental constraint.
\begin{table}
\begin{center}
\begin{minipage}[h]{0.47\textwidth}
      \centering
      \begin{tabular}{|c|c|c|c|}
      \hline\hline
      $\phi_i$ in $M_P$ & $n_s$ & $r$ & $ {N}$\\
      \hline
        28 &  0.9693 & 0.1633 & 47\\
        30 &  0.9733 & 0.1422 & 51\\
        32 &  0.9765 & 0.1250 & 54\\
        34 &  0.9792 & 0.1107 & 57\\
        36 &  0.9815 & 0.0988 & 59\\
        38 &  0.9833 & 0.0886 & 62\\
      \hline\hline
    \end{tabular}
     \caption{Data set for the inflationary parameters computed for quartic potential in view of \eqref{para4}, under the choice, ${\gamma V_1}= 0.0002,~ {V_0 \over V_1}=1 M_P^4$, $\alpha=0.5 M_P^2, ~\phi_f= 2.78198M_p$. The data deviates considerably from observational limit on `$r$' set up by Planck.}
      \label{tab:Table3}
   \end{minipage}
   \end{center}
\end{table}
Here again, for the sake of completeness, we compute the energy scale of inflation considering the relation \eqref{soln}, in view of the data, ($N = 51$, for which $\phi_i = 30M_P,~ \alpha=0.5M_P^2, ~{{V_0\over V_1}=1.0 M_P^4},~\gamma V_1= 0.0002$), appearing in Table 3. Accordingly, we find,
\begin{align}\label{ES5}{H_*}^2 = {11.316\over \gamma},\end{align}
which remains arbitrary unless $\gamma$ is fixed. The energy scale of inflation in a single scalar field model corresponding to GTR \cite{Wands}, as already considered earlier, is given by the following expression,

\begin{align}\label{ES6}{H_*} = 8\times 10^{13}\sqrt{r\over 0.2}GeV=6.744\times 10^{13}GeV =2.753\times 10^{-5} M_p,\end{align}
whose numerical value is calculated taking into account the value of the `tensor-to-scalar ratio $r = 0.1422$ from the same data set ($N = 51$) of Table-3. Thus, for $\gamma \approx 1.493 \times 10^{10}$,  the scale of inflation \eqref{ES5} is comparable with the single field scale of inflation \eqref{ES6}. Since $\gamma$ is fixed from physical consideration (sub-Planckian scale of inflation), the values of $\beta$, $V_1$ and $V_0$ may also be computed as,
\begin{align} \label{3V2}\beta = 1.382\times 10^8,\hspace{0.2 in} V_1= 1.340\times 10^{-14}, \hspace{0.2 in} V_0 = 1.0 V_1\approx 1.340\times 10^{-14}M_P^4.\end{align}
Finally, to display the graceful exit from inflation, we remind equation \eqref{S2}, which in view of the above form of the potential, $V(\phi)=V_0+{V_1\phi^4}$, is presented as,
\begin{align}\label{H3}-\frac{\gamma H^4}{V_1} - \frac{6\alpha H^2}{V_1}+\left[{\dot\phi^2\over 2V_1}+\left({V_0\over V_1}+{\phi^4}\right)\right]=0.\end{align}
During inflation, $H$ is slowly varying and all the terms in \eqref{H3} almost are of the same order of magnitude. However, as inflation ends the Hubble parameter decreases rapidly, and so both the left hand side terms ${\gamma H^4\over V_1}$ and ${\alpha H^2\over V_1}$ may be neglected without sacrificing generality, to find,
\begin{align}{\dot{\phi}}^2=-2\left[V_0+V_1\phi^4\right].\label{osc3}\end{align}
Taking into account, $\phi_f=2.78193M_P$, $V_0= 1.340\times 10^{-14}M_P^4$ and $V_1= 1.340\times 10^{-14}$ from Table-3, it is possible to demonstrate that the above equation indicates oscillatory behavior
\begin{align}\phi=\exp({i\omega t}),\end{align}
provided, $\omega \approx 4.594 \times 10^{-7}M_P$. Here again everything appears to be nice, except for the fact that the inflationary parameters obtained for the quartic potential do not quite agree with the observational constraint. The fact that quadratic and quartic potentials do not quite tally with observation, is no surprise, since Planck ruled out all simple integer power law potentials for single field inflation, strongly excluding the standard quadratic and quartic ones. In this context, the current analysis therefore agrees with Planck's observation. Next, we consider hilltop-like potentials with $V''(\phi) < 0$, which have been favoured by Planck \cite{Planck1, Planck2}.\\

\noindent
\textbf{Case-3:}\\
The results obtained form the quadratic and quartic potential suggest that, a flatter potential would be suitable and so we consider the third choice of the potential in the form, viz.,
\begin{eqnarray}\label{V1} V(\phi)= V_0-{V_1\over \phi},\end{eqnarray}
which remains almost flat initially for large value of the scalar field $\phi$ satisfying the condition $V''(\phi) < 0$. The expressions of $\epsilon, ~\eta$ \eqref{SR2} and $N$ \eqref{Nphi2} may then be found as,
\begin{align}\label{para2}\begin {split} &\epsilon =  \frac{\alpha}{\phi^4\left({V_0\over V_1}-{1\over \phi}\right)^2};\hspace{0.5 in}\eta =-\frac{4\alpha}{\phi^2(\frac{V_0}{V_1}\phi-1)};\\& N=\int_{\phi_f}^{\phi_i}\frac{3\phi^2\left[-3\alpha +\sqrt{9\alpha^2+\gamma V_1\left(\frac{V_0}{V_1}-\frac{1}{\phi}\right)}\right]}{\gamma V_1}d\phi.\end{split}\end{align}
The data set varying ${V_0 \over V_1}$ between $4.5M_P^{-1} \leq {V_0 \over V_1}\leq  5.2 M_P^{-1}$, displayed in Table-4, reveals that the agreement with observation is excellent with respect to the tensor to scalar ratio $r$ and the spectral index $n_s$. The number of e-folds lie within the range $40 \leq N \leq 47$, which is although a little low, but supposedly enough to resolve the horizon and flatness problems. Specifically, the last three data are very much acceptable. Note that the tensor to scalar ratio had been constrained from $r < 0.32$ to $r < .055$ in almost one decade. In any case, $r$ for the case-3 is capable of enduring further constraints, which might appear from future analysis.\\

\begin{table}
\begin{center}
\begin{minipage}[h]{0.49\textwidth}
      \centering
      \begin{tabular}{|c|c|c|c|c|}
     \hline\hline
      $\phi_f$ in $M_P$ &${V_0\over  V_1}$ in $M_P^{-1}$ & $n_s$ & $r$ & $ {N}$\\
      \hline
        0.5228 &4.5 &  0.9660 & 0.00496 & 40\\
        0.5156 &4.6 &  0.9668 & 0.00473 & 41\\
        0.5086 &4.7 &  0.9676 & 0.00452 & 42\\
        0.5019 &4.8 &  0.9684 & 0.00432 & 43\\
        0.4954 &4.9 &  0.9691 & 0.00413 & 44\\
        0.4891 &$5.0$ &  0.9698 & 0.00396 & 45\\
        0.4831 &$5.1$ &  0.9705 & 0.00380 & 46\\
        0.4772 &$5.2$ &  0.9711 & 0.00364 & 47\\
    \hline\hline
    \end{tabular}
     \caption{Data set for the inflationary parameters computed for equation \eqref{para2} with hilltop potential $V(\phi)= V_0-{V_1\over \phi}$, setting $\phi_i=3.1 M_P$, ${\gamma V_1}=.005 M_P^{5}$,~$\alpha=0.5 M_P^2$ and varying ${V_0 \over V_1}$. Agreement with Planck's observational data is outstanding.}
      \label{tab:Table4}
   \end{minipage}
   \end{center}
\end{table}
Next, let us compute the energy scale of inflation using the relation \eqref{soln}, selecting one of the data presented in Table 4, viz., ($N = 45$, for which $\phi_i = 3.1M_P,~ \alpha=0.5M_P^2, ~{{V_0\over V_1}=5.0 M_P^{-1}},~\gamma V_1= 0.005M_P^5$). Correspondingly we obtain,
\begin{align}\label{ES1}{H_*}^2 = {0.0078\over \gamma},\end{align}
which is still arbitrary, since $\gamma$ has not yet been fixed. Now, the energy scale of inflation for a single scalar field model corresponding to GTR, as already considered earlier, is represented by the following relation \cite{Wands},
\begin{align}\label{ES2}{H_*} = 8\times 10^{13}\sqrt{r\over 0.2}~GeV=1.128\times 10^{13}~GeV =0.4604\times 10^{-5} M_p.\end{align}
In the above, we computed the numerical value taking into account the tensor-to-scalar ratio $r = 0.00396$ corresponding to $N = 45$, presented in Table 4. In reverse for a cross check, let us consider that inflation lasted for $\Delta t \approx 10^{-36}~s$, if it is assumed to occurs between $(10^{-40}$ to $10^{-36})~s$. Then for the particular value of $N = 45$, which is precisely $\Delta N$, $H \approx {45\over 10^{-36}}~s^{-1}$. However, in our chosen system of unit, $1 s^{-1} \approx {1\over 3.65 \times 10^{42}}~M_P$. Hence, $H = H_* \approx 1.23 \times 10^{-5}~M_P$. Thus, to arrive at this particular scale of inflation, we need to restrict $\gamma \approx 3.68 \times 10^{8}$. Once the parameter $\gamma$ is fixed from physical consideration (sub-Planckian scale of inflation), the values of $\beta$, $V_1$ and $V_0$ are found as well,
\begin{align} \label{V2} \beta = {\gamma\over 108} = 3.41\times 10^6,\hspace{0.2 in} V_1 \approx 1.35\times 10^{-11}M_P^5, \hspace{0.2 in} V_0 = 5.0 V_1\approx 6.75\times 10^{-11}M_P^4.\end{align}
Finally, it is now required to check for graceful exit from inflation in the present model. We therefore recall equation \eqref{S2}, which for the potential, $V(\phi)=V_0-{V_1\over \phi}$, under consideration, is expressed as,
\begin{align}\label{H1}-\frac{\gamma H^4}{V_1} - \frac{6\alpha H^2}{V_1}+\left[{\dot\phi^2\over 2V_1}+\left({V_0\over V_1}-{1\over {\phi}}\right)\right]=0.\end{align}
Initially, during inflation all the terms in \eqref{H1} are more-or-less of the same order of magnitude, since $H$ is slowly varying. Nonetheless as inflation halts, the Hubble parameter falls of sharply and so one can neglect both the terms ${\gamma H^4\over V_1}$ and ${\alpha H^2\over V_1}$ appearing in the left hand side of \eqref{H1} without any loss of the generality. Finally one arrives at,
\begin{align}{\dot{\phi}}^2=-2\left[V_0-{V_1\over \phi}\right] .\label{osc2}\end{align}
Taking into account, $\phi_f=0.4891M_P$, $V_0= 6.75\times 10^{-11}M_P^4$ and $V_1= 1.35\times 10^{-11}M_P^5$ from Table-4, one can trivially exhibit the following oscillatory behavior of the scalar field at the end of inflation,
\begin{align}\phi=\exp({i\omega t}),\end{align}
provided, $\omega \approx  1.826\times10^{-5}M_P$, and hence graceful exit from inflation is possible. In a nutshell, the potential \eqref{V1} is well-suited in the very early cosmic evolution. \\

\noindent
\textbf{Case-4:}\\
Finally, we consider yet another form of the potential, viz.,
\begin{eqnarray} \label{V11} V(\phi)= V_0-{V_1\exp(-b\phi)},\end{eqnarray}
which again satisfies the condition $V''(\phi) < 0$. Such a potential had been widely used to study inflation because of having excellent feature. The almost flat behaviour of the potential at the advent of inflation, when $\phi$ is large enough, allows slow roll to be conceivable, while as $\phi$ decays, it leaves a bare cosmological constant behind. The expressions of $\epsilon, ~\eta$ \eqref{SR2} and $N$ \eqref{Nphi2} are now found as,
\begin{align}\label{para3}\begin {split} &\epsilon= \alpha \bigg({V'\over V}\bigg)=\frac{\alpha b^2\left[\exp(-b\phi)\right]^2}{\left[\frac{V_0}{V_1}-\exp(-b\phi)\right]^2} ;\hspace{0.5 in}\eta =-\frac{2\alpha b^2\exp(-b\phi)}{\left[\frac{V_0}{V_1}-\exp(-b\phi)\right]};\\& N=\int_{\phi_f}^{\phi_i}\frac{3\left[-3\alpha +\sqrt{9\alpha^2+\gamma V_1\left(\frac{V_0}{V_1}-\exp(-b\phi)\right)}\right]}{\gamma V_1b\exp(-b\phi)}d\phi.\end{split}\end{align}
In Table-5, we exhibit a data-set corresponding to the expressions \eqref{para3}, varying ${V_0 \over V_1}$ between $1.194 \leq {V_0 \over V_1} \leq 1.212$. The agreement of $r$ and $n_s$ with observation is excellent, while the variation of the number of e-folds $(71 \leq N \leq 73)$, is not too large. The scalar to tensor ratio can confront further constraints, which as mentioned might appear from future analysis.
\begin{table}
\begin{center}
\begin{minipage}[h]{0.47\textwidth}
      \centering
      \begin{tabular}{|c|c|c|c|c|}
     \hline\hline
      $\phi_f$ in $M_P$ &${V_0 \over V_1}$  & $n_s$ & $r$ & $ {N}$\\
      \hline

        0.05024 &1.194 &  0.9601 & 0.02215 & 71\\
        0.04466 &1.196 &  0.9602 & 0.02206 & 71\\
        0.03909 &1.198 &  0.9603 & 0.02197 & 72\\
        0.03353 &1.200 &  0.9604 & 0.02189 & 72\\
        0.00587 &1.210 &  0.9609 & 0.02147 & 73\\
        0.00036 &1.212 &  0.9610 & 0.02138 & 73\\

    \hline\hline
    \end{tabular}
     \caption{Data set \eqref{para3} for the inflationary parameters computed using \eqref{para3} with the potential $V(\phi)= V_0-{V_1\exp(-b\phi)}$ setting $\phi_i=5.75 M_P$, ${\gamma V_1}=7.0 M_P^{4}$,~$b=0.3M_P^{-1}, ~\alpha=0.5 M_P^2$ and under the variation of ${V_0 \over V_1}$. Here again the data set lies appreciably within the observation limit set up by Planck.}
      \label{tab:Table5}
   \end{minipage}
   \end{center}
\end{table}
Let us also compute the energy scale of inflation considering the relation \eqref{soln}, in view of the data, ($N = 72$, for which $\phi_i = 5.75M_P,~ \alpha=0.5M_P^2, ~{{V_0\over V_1}=1.2},~\gamma V_1= 7.0M_P^4$), appearing in Table 5. Correspondingly, we compute,
\begin{align}\label{ES3}{H_*}^2 = {1.567\over \gamma},\end{align}
which, as before remains arbitrary unless $\gamma$ is fixed. The energy scale of inflation in a single scalar field model corresponding to GTR \cite{Wands}, as already considered earlier, is presented by the following expression,
\begin{align}\label{ES4}{H_*} = 8\times 10^{13}\sqrt{r\over 0.2}GeV=2.647\times 10^{13}GeV =1.080\times 10^{-5} M_p,\end{align}
whose numerical value is calculated taking into account the value of the tensor-to-scalar ratio $r = 0.02189$ from the same data set ($N = 72$) of Table-5 under consideration. Again as before taking $\Delta t \approx 10^{-36}~s$ and $\Delta N = 72$, one can compute $H = H_* \approx 1.97\times 10^{-5}~M_P$. Thus, for $\gamma \approx 1.343 \times 10^{10}$ the scale of inflation \eqref{ES3} is comparable with the single field scale of inflation \eqref{ES4}. Since $\gamma$ is fixed from physical consideration (sub-Planckian scale of inflation), the values of $\beta$, $V_1$ and $V_0$ may also be computed, which are,
\begin{align} \label{2V2}\beta = 1.243\times 10^8,\hspace{0.2 in} V_1= 5.212\times 10^{-10}M_P^4, \hspace{0.2 in} V_0 = 1.2 V_1\approx 6.254\times 10^{-10}M_P^4.\end{align}
Finally, to exhibit graceful exit from inflation, we use equation \eqref{S2}, which in view of the above form of the potential, $V(\phi)=V_0-{V_1\exp(-b\phi)}$, is expressed as,
\begin{align}\label{H2}\frac{\gamma H^4}{V_1} + \frac{6\alpha H^2}{V_1} - \left[{\dot\phi^2\over 2V_1}+\left({V_0\over V_1}-{\exp(-b\phi)}\right)\right]=0.\end{align}
During inflation, $H$ is slowly varying and all the terms in \eqref{H2} almost are of the same order of magnitude. However, as inflation ends the Hubble parameter decreases rapidly, and so both the left hand side terms ${\gamma H^4\over V_1}$ and ${\alpha H^2\over V_1}$ may be neglected without loss of generality, to find,
\begin{align} {\dot{\phi}}^2=-2\left[V_0-V_1\exp(-b\phi)\right].\label{osc3}\end{align}
Taking into account, $\phi_f=0.03353M_P$, $V_0= 6.254\times 10^{-10}M_P^4$ and $V_1= 5.212\times 10^{-10}M_P^4$ from Table-5, it is possible to demonstrate that the above equation exhibits oscillatory behavior
\begin{align}\phi=\exp({i\omega t}),\end{align}
provided, $\omega \approx 4.412 \times 10^{-4}M_P$.\\

In a nut-shell, the inflationary parameters computed for the form $f(X) = \alpha X - \beta X^2, ~\alpha,\beta > 0$, associated with the metric as well as symmetric teleparallel gravity theories in scalar field driven inflation, are in excellent agreement with the observed inflationary parameters, for both the third and fourth choices of flat potentials $V(\phi) = V_0 - {V_1\over \phi}$ and $V(\phi) = V_0 -V_1e^{-b\phi}$, which satisfy the condition $V''(\phi) < 0$, as suggested by Planck. It is worth mentioning that Planck's observation using 6-parameter $\Lambda$CDM model \cite{Planck1,Planck2} overruled power law (quadratic, quartic etc.) potentials due to the constraint on tensor to scalar ratio.In recent years, ACT also strongly favoured slow-roll single field inflation with a flat (plateau type) potential. Indeed we also find that the inflationary parameters deviate from experimental values for the first and second choices of quadratic and quartic potentials $V(\phi) = V_0 + {V_1 \phi^2}$ and $V(\phi) = V_0 +V_1\phi^4$ respectively, since $r$ is considerably large.

\subsection{Reheating:} 
As inflation is on the verge of termination $\epsilon \approx 1$, a scalar field $\chi$ originating from quantum fluctuation but remained subdominant during inflationary epoch, interacts with the inflaton field $\phi$ such that the background equation reads as,
\begin{equation}\label{P1} S = \int d^4x\sqrt{-g}\left[{1\over 2}\phi_{,\mu}\phi^{,\mu} - V(\phi) + {1\over 2}\chi_{,\mu}\chi^{,\mu} -{1\over 2}w^2\phi^2\chi^2\right]\end{equation}
where $w$ is a coupling constant. As mentioned, both Planck and ACT favoured flat plateau-type potential and indeed we also find that the potential $V(\phi)= V_0-{V_1\exp(-b\phi)}$ agrees with the observational constraints fairly well. Hence we consider it for exhibiting reheating. The standard preheating phenomena requires that at the end of inflation the inflaton field should oscillate at the minimum of the potential which happens for power law potentials. The problem with the flat potential is $V' = b_1 e^{-b\phi}$ being positive definite, it decreases monotonically, having no finite minimum. Therefore, there are no natural harmonic oscillations too and standard preheating due to harmonic oscillation of the inflaton remains obscure. Nonetheless, particles (bosons or fermions) may be produced almost instantly through the violation of adiabaticity as the inflaton field rolls rapidly towards smaller value. The adiabaticity breaks down at $|\dot\phi_*| \ge w\phi^2$ where $\phi_*$ is the point of maximal non-adiabaticity. For interaction type $\mathcal{L}_{int} = {1\over 2}w^2 (\phi-\phi_*)^2\chi^2$, the inflaton field while rapidly crossing $\phi_*$, changes the effective interaction scalar mass $m_\chi = w|\phi - \phi_*|$. As a result, huge number of particles are produced almost instantaneously, whose number density is $n_\chi \approx {\left(w |\dot\phi_*|\right)^{3\over 2}\over 8\pi^3}$. This mechanism is called `instant (non-oscillatory) preheating mechanism' \cite{P1}. As $\chi$ decays to lighter particles, reheating becomes effective. \\

During reheating, the inflaton field decays and transfers energy to a thermal bath of radiation which initiates hot Big-Bang. This energy transfer is modelled by adding a decay width parameter ($\Gamma$). The Klein-Gordon equation is therefore modified in the presence of the friction term ($\dot\phi\Gamma$) in addition to already existing $3H\dot\phi$ term as, 
\begin{align} \ddot{\phi}+\left(3H+\Gamma\right)\dot\phi + bV_1 e^{-b\phi}= \dot{\rho_{\phi}}+3H(\rho_{\phi}+p_{\phi})+\Gamma{\dot\phi}^2=0,\end{align}
where $\rho_{\phi}={1\over 2}{\dot\phi}^2+ V(\phi)$ and $p_{\phi}={1 \over 2}{\dot\phi}^2 - V(\phi)$. The radiation production equations is,
\begin{align}\label{ET}
\dot{\rho_{r}}+4H\rho_{r}=\Gamma{\dot\phi}^2,\end{align}
where $\rho_r$ stands for radiation energy density. Since in the early universe quadratic correction term dominates, so $108 \beta H^4 \gg 6\alpha H^2$ and hence $H \propto (\rho_\phi)^{1 \over 4}$, instead of $(\rho_\phi)^{1 \over 2}$ which appears in the standard FLRW model. This effectively changes reheating duration, damping of inflaton motion, radiation dilution, and reheating temperature altogether. Now, reheating completes and ends when the Hubble rate becomes comparable to the decay width
\begin{align} H_{reh}\approx \Gamma.\end{align}
At that stage, the universe is radiation dominated and so,
\begin{align}\rho_r(t_{reh})\approx \rho_{tot}(t_{reh}).\end{align}
Hence the Friedmann equation \eqref{S1} now reads as, 
\begin{align} \label{Re}\rho_r(t_{reh})= 6\alpha \Gamma^2 + 108\beta \Gamma^4 \end{align}
where we set $H = \Gamma$. At this stage, universe is a thermal bath containing $g^*$ effective relativistic degrees of freedom and so we have,
\begin{align}\rho_r = {\pi^2\over 30}g^*{T_{reh}}^4.\end{align}
Finally, in view of \eqref{Re}, the expression for reheating temperature reads as, 
\begin{align}\label{treh} T_{reh}=\left[{30 \over \pi^2g^*}(6\alpha \Gamma^2 + 108\beta \Gamma^4 )\right]^{1 \over 4}.\end{align} 
One an easily check that setting $\beta \approx 0$, the above equation takes the form
\begin{align} T_{reh}= \left[{30 \over \pi^2g^*}(6\alpha \Gamma^2)\right]^{1 \over 4} = \left[{90\over \pi^2 g^*}\right]^{1\over 4} \sqrt{\Gamma M_p}, \end{align}
where we have replaced $\alpha = {M_P^2\over 2}$ to recover the standard Friedmann equation. Now since the life-time of gravitinos, which interact only with gravity, are extraordinarily high and therefore their energy decay product might ruin predictions of BBN, so it is customary to take $\Gamma= 10^5 ~Gev$ to avoid such problem. It also satisfies Higgs' vacuum stability limit and serves as a benchmark of low-sale leptogenesis. Further choosing $g^*=100$, $\beta = 1.243 \times 10^8$ \eqref{2V2}, $\alpha = 0.5M_P^2$ where $M_p = 2.435\times 10^{18} Gev$, the reheating temperature \eqref{treh} is found to be,
\begin{align} T_{reh}=\left[{30 \over 100\pi^2}(1.779\times 10^{47} + 1.342\times 10^{30})\right]^{1 \over 4}Gev \approx 2.7118\times 10^{11}~Gev,\end{align}
which is $10^{14}$ times higher than the temperature ($1~Mev$) at BBN. In fact, even a very small decay width $\Gamma \approx 10^{-5}$ leads to reheating temperature $T_{reh} \approx 86~Mev$. To be precise, the $f(X)$ model under present  consideration gives way to the hot Big-Bang following a scalar field driven inflation and reheating. Our next task is therefore to inspect the viability of the model in the radiation dominated era.

\section{Radiation and early Matter dominated eras:}

Exhibiting a viable inflationary era in the early universe and unifying it with late-time cosmic accelerated expansion is not enough. At the end of inflation, particles are produced giving way to the hot big-bang. Thereafter the universe enters the radiation dominated era following appropriate reheating as depicted above. Subsequently it enters the early matter dominated era These to eras are best described by the Friedmann-Lema$\hat{\mathrm{i}}$tre-like standard model solution $a\propto \sqrt t$ and $a\propto t^{2\over 3}$; respectively, a decelerated expansion is amenable though. Let us therefore check if such a viable radiation $(p ={1\over3} \rho)$ and early matter dominated $(p = 0)$ eras are admissible with the torsion induced or non-metricity induced theory of gravity. Here, $p$ stands for the thermodynamic pressure while $\rho$ is the matter density of the perfect fluid together with the dark matter.\\

As mentioned, in the context of generalized symmetric teleparallel gravity, it is suggested that a term $-6\lambda M^2\left({Q\over 6 M^2}\right)^\delta$ solves the cosmic puzzle without dark energy, for $\delta < 0$. The term $M$ is included as a scale, so that as the Hubble parameter $H > M$, inflation is realized in the very early universe, while as $H < M$, late-time acceleration \cite{coincident} is admissible. However, we have noticed that vacuum teleparallel equations do not admit such a form of $f(X)$ unless an additional field, such as a scalar field is incorporated. The fact that a scalar field necessarily drives successful inflation in the present context, has also been explored. The scale $M$ is therefore absolutely unnecessary, since any additional term should only be effective at the late-stage of cosmic evolution. We therefore simply consider an additional term such as ${\eta\over\sqrt{-X}}$, setting $\delta = -{1\over 2}$ for simplicity to inspect the consequence. The generalized form is now $f(X)= \alpha_1 X+ \beta_1 X^2+{\eta\over {\sqrt{-X}}}$, where $\eta$ is a constant. The field equations incorporating barotropic fluid may therefore be written as,
\begin{eqnarray}\label{L1}\begin {split}& 6\alpha H^2+108\beta H^4+{2\eta\over {\sqrt6} H} = \rho,\\&
6\alpha H^2+108\beta H^4+ 4\alpha \dot H + 144\beta H^2 \dot H +{2\eta\over {3\sqrt 6} H}\left(3 - {\dot H\over H^2}\right) = - p.\end{split}\end{eqnarray}
One can easily notice that under the choice $a = a_0 t^n$, the first field equation of \eqref{L1} reads as
\begin{eqnarray} \label{eta} 6\alpha {n^2\over t^2} + 108 \beta{n^4\over t^4} + {2\eta\over \sqrt{6} n}t = \rho.\end{eqnarray}
The onset of the radiation era is at around $t \approx 10^{-26}~ sec$, when the last term in \eqref{eta} remains subdominant while the second term dominates over others. Since conservation of energy-momentum tensor of barotropic fluid demands $\rho a^4 = \rho_{r0}$, therefore $\rho = {\rho_{r0}\over a_0^4 t^2}$ doesn't satisfy the above field equation. To explicitly see the consequence, we initially disregard the last term $(\eta = 0)$, substitute $(p = {1\over 3}\rho)$ and eliminate $\rho$ from the pair of equations \eqref{L1}, to find\\
\begin{eqnarray} 24\alpha H^2+12\alpha\dot H+432\beta H^2\dot H+432\beta H^4=0~~~\rightarrow~~~\left[\frac{\alpha+36\beta H^2}{(2\alpha+36\beta H^2)H^2}\right]dH=-dt.\end{eqnarray}
The above equation, upon integration gives,
\begin{eqnarray}\label{tH} t=\frac{1}{2 H}-\frac{3 \sqrt{\beta } \tan ^{-1}\left(\frac{3 \sqrt{2\beta}  H}{\sqrt{\alpha }}\right)}{\sqrt{2\alpha}},\end{eqnarray}
which indicates that for expanding universe $H > 0$, the second term dominates over the first, resulting in a negative $t$. On the contrary, for positive $t$ a negative $H$, i.e contracting model results. Consequently, a viable (decelerated) radiation era in this model starts at a later stage $t > 1$, as the second term becomes subdominant. However, it doesn't last long since the last term starts dominating soon as $t\gg 1~sec$. In the matter-dominated era $(p = 0,~\rho = {\rho_{m0}\over a^3} ={\rho_{m0}\over a_0^3 t^2})$, the third term dominates from the very beginning and the field equations \eqref{eta} are never satisfied. Hence, associating the last term ($\eta \ne 0$) for late-time cosmic acceleration tells upon both the radiation as well as on the matter (pressureless-dust) dominated eras. On the contrary, in the absence of the last term, although Radiation era starts a little late $t > 1~sec$, it continues and gives way to the Friedmann-like matter dominated era. But then, solution of the cosmic puzzle requires to incorporate additional field. In a nut-shell, unlike modified theories of gravity, teleparallel gravity theories cannot induce a viable cosmological evolution from geometry alone and somehow an additional field is required for the purpose. \\

The only way to circumvent the associated problem is to incorporate the scalar field, however small, assuming not all the scalar has been used up in creating particles after inflation terminates. Disregarding the third term $(\eta = 0)$, equations \eqref{L1} now reads as,
\begin{eqnarray}\begin{split}\label{rhop} &6\alpha H^2+108\beta H^4=\rho+{1\over 2}{\dot\phi}^2+V(\phi), \\& 6\alpha H^2+4\alpha \dot {H}+144\beta\dot{H}H^2+108\beta H^4=-p-{1\over 2}{\dot\phi}^2+V(\phi).\end{split}\end{eqnarray}
Combining the above pair of equations for radiation $(p = {1\over 3} \rho)$, one obtains,
\begin{eqnarray}\label{comR} {\dot\phi}^2=-12(\alpha+36\beta H^2)\dot H-12(2\alpha+36\beta H^2)H^2+4V(\phi).\end{eqnarray}
For a power law solution in the form $a=a_0 t^n$, equation \eqref{comR} may be expressed as,
\begin{eqnarray} \label{phirad}{{\dot\phi}^2\over 12}= \frac{\alpha n}{t^2}+\frac{36\beta n^3}{t^4} -\frac{2\alpha n^2}{t^2}-\frac{36\beta n^4}{t^4}+{V(\phi)\over 3}.\end{eqnarray}
For a viable (Friedmann-like) radiation-dominated era, if we choose $n = {1\over 2}$, then \eqref{phirad} finally takes the following form,
\begin{eqnarray}\label{rad} {{\dot\phi}^2\over 12}= \frac{9\beta}{4t^4} +{V(\phi)\over 3}.\end{eqnarray}
Now, for a decaying scalar field $\phi=\phi_0 t^{-m}$ (say) the above equation \eqref{rad} reduces to,
\begin{eqnarray} { m^2{\phi_0}^2\over 12 t^{2(m+1)}}-{9\beta\over 4t^4}-{V(\phi)\over 3}=0,\end{eqnarray}
and is satisfied exactly for $m = 1$, i.e.,
\begin{eqnarray}\label{PHI}\phi = {\phi_0\over t};~~~V(\phi) = V_0\phi^4; ~~~V_0=\frac{1}{4\phi_0^2}-\frac{27\beta}{4\phi_0^4}.\end{eqnarray}
Undoubtedly, this is a nice result, since now the pure radiation era is initiated even in the presence of the second term, just after the end of inflation and continues giving way to the matter dominated era after successful formation of the CMBR, when photon decouples and redshifted. In the matter dominated era $(p = 0)$, the field equations may be recast as,
\begin{eqnarray}\label{m}\begin{split}& 2\alpha\left(2{\ddot a\over a} + {\dot a^2\over a^2}\right) = -108\beta{\dot a^4\over a^4}-144\beta{\dot a^2\over a^2}({\ddot a\over a}-{\dot a^2\over a^2}) - {1\over 2}\dot\phi^2 + V(\phi) =-p_{eff}\\&
6\alpha{\dot a^2\over a^2} = \rho -108\beta{\dot a^4\over a^4} + {1\over 2}\dot\phi^2 + V(\phi)  = \rho_{eff}\end{split}. \end{eqnarray}
Hence the effective state parameter $\omega$ reads as,
\begin{eqnarray}\label{omega} \omega = {36\beta {\dot a^2\over a^2}(4{\ddot a\over a}-{\dot a^2\over a^2}) + {1\over 2}\dot\phi^2 - V(\phi) \over \rho -108\beta{\dot a^4\over a^4} + {1\over 2}\dot\phi^2 + V(\phi)} = {36\beta H^2(4\dot H+3H^2) + {1\over 2}\dot\phi^2 - V(\phi) \over \rho -108\beta H^4 + {1\over 2}\dot\phi^2 + V(\phi)}.\end{eqnarray}
Clearly, if the potential $V(\phi) = V_0 \phi^4$ dominates at the later epoch of cosmic evolution, then acceleration might follow early decelerated expansion. This is possible, if $\phi_0$ is very small so that ${\beta\over{\phi_0}^4}$ term appearing with negative sign in $V_0$ \eqref{PHI} dominates over others and the denominator becomes negative at the late-stage of cosmic evolution. But then, quartic potential is ruled out by the Planck's data. So, let us consider the third case \eqref{V1} having the potential $V(\phi)= V_0-{V_1\over \phi}$, which showed considerably good agreement with observations. The equation \eqref{rad} in the Friedmann-like radiation dominated era ($p = {1\over 3}\rho,~a\propto t^{1\over 2}$) now takes the form,
\begin{eqnarray}\label{sol}{{\dot\phi}^2\over 12}= {9\beta \over 4t^4}+{1\over 3}\left[V_0-{V_1\over \phi}\right],\end{eqnarray}
which can not be solved analytically. Thus, let us try to solve it numerically by substituting the values of $V_0,~V_1, ~\beta$ in view of \eqref{V2} fixed at the time of inflation, in equation \eqref{sol} to find,
\begin{eqnarray}\label{Rad}{{\dot\phi}^2}- {22.4\times 10^{-9}}+{{3.45\times 10^{-10}}\over \phi}-{{2.89\times 10^8 }\over t^4}= 0.\end{eqnarray}
If we consider that the pure radiation era initiated at $10^{-26}~\mathrm{s}$ (when the value of the scalar field was as small as $\phi \approx 10^{-10}~M_P$), and lasted till $4,00,000 ~\mathrm{yrs} \approx 10^{13}~\mathrm{s}$, then the plot $\phi$ versus $t$ depicted in figure-5, shows enormous increase of $\phi_{MRE} \approx 10^{68}~M_P$ (MRE in the suffix stands for Matter-Radiation-Equality) at the onset of matter-dominated era. Thus, the matter dominated era accompanies such a huge amount of the scalar field. However, figure-6 depicts that scalar field remains constant at pressure-less dust-dominated era. Since the form of $\phi$ is not known, so it is not possible to solve the first equation \eqref{rhop}. 
\begin{figure}
\begin{minipage}[h]{0.47\textwidth}
\centering
\includegraphics[width=1.0\textwidth] {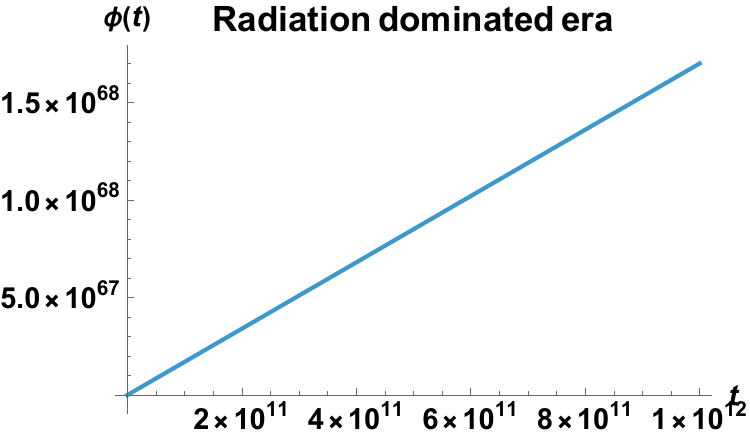}
 \caption{The plot $\phi~ \mathrm{vs}~ t$ for positive root of $\phi$, taking $\phi_{\mathrm{initial}} = 10^{-10}~M_P$ at $t_{\mathrm{initial}} = 10^{-26}~s$. Clearly $\phi$ increases exorbitantly to $\phi_{MRE} = 10^{68}~M_P$ at Matter-Radiation Equality (MRE).}
      \label{fig:5}
   \end{minipage}%
  \hfill
\begin{minipage}[h]{0.47\textwidth}
\centering
\includegraphics[ width=1.0\textwidth] {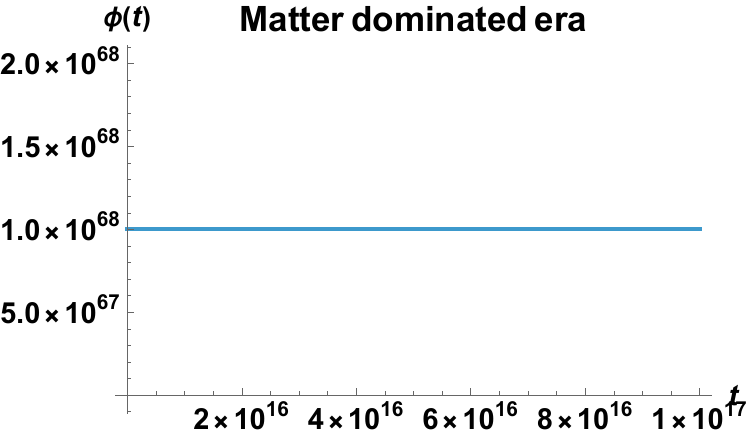}
 \caption{The plot $\phi~ \mathrm{vs}~ t$ for positive root of $\phi$, taking $\phi_{\mathrm{MRE}} = 10^{68}~M_P$ at  Matter-Radiation Equality. Note that $\phi$ remains constant till the onset of accelerated expansion $z \approx 1$, i.e., $t \approx 5~Gyr \approx 10^{17}~s$.}
      \label{fig:6}
   \end{minipage}%
\end{figure}

Likewise, in the case of exponentially decaying potential \eqref{V11} the equation is expressed as,
\begin{eqnarray}\label{ep1} \dot \phi^2 = {27\beta\over t^4} + {4 V_0\over V_1} -4(1-b \phi),\end{eqnarray}
where we have neglected higher order terms in the exponential part, since we assume the presence of a small amount of scalar field $\phi \approx 10^{-10}~M_P$ at the beginning of the radiation dominated era as before, while $b = 0.3~M_P^{-1}$ (see table 5). Consequently, after substituting $V_0 = 6.254\times 10^{-10}~M_P^4$ and $V_1 = 5.212\times 10^{-10}~M_P^4$, the equation \eqref{ep1} reads as,
\begin{eqnarray}\label{ep2} \dot \phi^2 = 0.8 + 1.2 \phi + {3.356\times 10^{9}\over t^4}.\end{eqnarray}
Although, equation \eqref{ep2} looks quite different from \eqref{Rad}, but it also exhibits identical feature as figure 5 and thereafter remains constant in the matter dominated era as depicted in figure 6 when plotted and so are not displayed. In a nut-shell, the form $f(X) = \alpha X - \beta X^2, ~\alpha,\beta > 0$, associated with the generalized metric as well as general symmetric teleparallel gravity theories for quartic scalar field potential yields exactly the Friedmann-Lema$\hat{\mathrm{i}}$tre like radiation era in the presence of a scalar field, which decays with cosmic evolution. But then, quartic potential is ruled out by observation from Planck and ACT, as we have also noticed and it is true for the quadratic potential as well. On the contrary, both the choice of the potentials $V(\phi) = V_0 - {V_1\over \phi}$ and $V(\phi) = V_0 -V_1e^{-b\phi}$ show excellent agreement with the said observation and also admit a Friedmann-Lema$\hat{\mathrm{i}}$tre-like radiation dominated era. However, the scalar field increases exorbitantly leaving its trace as dark energy in the late universe. Thus, late-time acceleration becomes an artefact of the scalar field leaving teleparallel theories in awe.

\section{Covariant scalar-vector-tensor formalism:}
So far, we have dealt with pure geometric point of view supplementing the action with a minimally coupled scalar field. Nonetheless, covariant actions for scalar-torsion and scalar-vector-tensor non-metricity theories exist, which are presented in \eqref{Tcov1} and \eqref{Qcov1} respectively. In the metric teleparallel gravity theory in particular, the LLI action is presented in the most general form \eqref{Tcov1}, while the connection variation equation \eqref{con} is satisfied trivially for $(k = 0 ,\pm 1)$ in the RW metric \eqref{RW}. It may be noted that the scalar field must be retained throughout the cosmic evolution for the action not to crash and hence the issue of dark energy cannot be avoided. Additionally, action \eqref{Tcov1} may be reduced to the Einstein's frame under appropriate conformal transformation, so that it takes the form we dealt with minimally coupled scalar field. Thus, all results and the discussions made so far are valid for covariant scalar-tensor metric teleparallel gravity also. \\

In contrast, the covariant GSTG action may be expressed in scalar-vector-tensor form \eqref{Qcov1}, and one can use the affine connections \eqref{Hoh11} to compute the following term \cite{DA},
\begin{eqnarray}\label{Q2}\left(Q^\alpha - \hat{Q}^\alpha\right)f_{,\alpha}(\Phi) = - {6 \dot a\over a}f'{\Phi}\dot \Phi.\end{eqnarray}
In the above, we have made the choice $\mathrm{N} = 1$ because diffeomorphic invariance $\mathcal{H} = \mathrm{N}\mathrm{H}$, has been established \cite{DA}, where $\mathrm{H}$ is the Hamiltonian. The action therefore reads as,
\begin{eqnarray}\begin{split} \label{A2} A_{\mathrm{cov}1}&=\int\left[ -3a\dot a^2f(\Phi)-3a^2 \dot a f'{\dot\Phi}+\left(\omega(\Phi) {\dot\Phi}^2 - U(\Phi)\right) a^3\right]dt,\end{split}\end{eqnarray}
The above action simply turns out to be the `non-minimally coupled scalar-tensor theory of gravity' and therefore undoubtedly LLI. This has been widely explored in the context of early as well as the late universe over decades, since the advent of Brans-Dicke theory. Here again the scalar must not vanish at any stage, as is evident from the action and therefore retain its very presence as dark energy. \\

To be very precise, we show that without a scalar field both the generalized metric teleparallel gravity theory (GMTG) and the generalized symmetric teleparallel gravity theory (GSTG) are unable to explain early as well as the late stages of cosmic evolutions in the flat FLRW metric.

\section{\bf{Concluding remarks:}}
In this article, we emphasized that in the flat FLRW background \eqref{RW}, the torsion scalar takes the form $\mathrm{T} = -6H^2$ both in covariant (in spherical coordinates including flat spin connection) and non-covariant (in diagonal Cartesian coordinate, being devoid of spin connection) formalisms and show up identical field equations, since spin connection appears only as a gauge. Further, out of three possible forms of $Q$ obtainable in view of different gauge choices in flat ($k = 0$) RW metric for covariant symmetric teleparallel theory, only one can be generalized to $f(Q), Q\ne 0$ theory \cite{DA2}, which again results in $Q = -6H^2$, while connection variation equation is identically satisfied. It may be mentioned that in coincidence gauge ($\Gamma = 0$) too the same form of $Q = -6H^2$ is realizable and so dynamics remain unchanged. Since field equations of symmetric teleparallel gravity remain unaltered from the metric teleparallel gravity, so we have treated both on the same footing considering $Q = \mathrm{T} = X = -6H^2$ without any loss of generality. Thus, our analysis holds for GMTG and GSTG in general, in the flat FLRW background metric. These teleparallel gravity theories are already known to suffer from some serious pathologies, which we discuss in the introduction to some details. In the present article, we closely inspect some additional features in respect of credentials and illegitimacy of both the teleparallel gravity theories.\\

Firstly, we show that since pure vacuum solution doesn't admit arbitrary form of $f(X)$, so unlike curvature induced Starobinsky inflation, torsion/non-metriity induced inflation is unrealisable. We therefore have chosen a specific form  $f(X) = \alpha X - \beta X^2$, with $\alpha > 0, \beta > 0$ as a minimal generalization over TEGR/STEGR, being established through reconstruction programme where the sign of the coupling parameters have been fixed in view of energy conditions. We then re-examine the issue of 'Branched Hamiltonian'. While earlier, taking into account $f(X) = \alpha X + \beta X^2$, with $\alpha > 0, \beta > 0$ arbitrarily, the theory was found to suffer from the issue of `Branched Hamiltonian', the issue doesn't exist for the present form any more. This is a promising outcome, since the issue has no practical resolution. Next, we study a minimally coupled scalar field-driven inflation empowered with four different (power law and flat) potentials $V(\phi) = V_0+V_1\phi^2,~V(\phi) = V_0+V_1 \phi^4,~V(\phi) =  V_0 - {V_1\over \phi}$ and $V(\phi) =  V_0 - V_1 e^{-b\phi}$. The inflationary parameters, viz., the tensor to scalar ratio ($r$) and the spectral index ($n_s$) are found to agree quite appreciably with the latest released Planck and ACT datasets for the last two flat potentials, under the standard choice $\alpha = {1\over 2}~ M_P^2$ (to recover GTR and Poisson equation in appropriate limits), while it is ruled out for the quadratic and quartic potentials. The parameter $\beta$ is fixed comparing the energy scale of inflation with that of the single-field inflation in GTR and the theory is found to exit gracefully from inflation. We also study reheating, which exhibits appropriate reheating temperature and gives way to hot Big-Bang, since quartic term doesn't really contribute. This is yet another nice result. Since the universe enters hot Big-Bang phase, we next search for a viable radiation era. It has been clearly shown that additional term such as $X^{-\eta},~\eta > 0$ suggested earlier for late time cosmic acceleration, no way can provide a viable radiation era, early matter dominated era and cause accelerated expansion in the late-stage of matter dominated era, without pathological behaviour, such as increase of the ordinary matter-density. In principle, it is revealed that in the absence of a scalar field, a viable (Friedmann-like) radiation era is inconceivable. On the other hand, considering the presence of the scalar field, however small, Friedmann-like radiation era is attainable for quartic potential only, which is ruled out clearly by Planck and ACT observations. In contrast, both the flat third and the fourth potentials admit FLRW-like radiation era $a(t) \propto \sqrt t$, but the scalar field increases exorbitantly, keeping its trace as dark energy in the matter-dominated era. Finally, the study with covariant (scalar tensor and scalar-vector-tensor) formalisms also depict identical features. In a nut-shell, both the GMTG ($f(\mathrm{T})$ theory) and GSTG ($f(Q)$ theory) are incompatible to expatiate any stage of the history of cosmic evolution without associating an extra field in the FLRW background. This goes against the very first motivation for introducing these theories. Finally we discussed that no new result is expected from scalar-tensor equivalent GMTG theory. However, the scalar-vector-tensor formalism of symmetric teleparallel gravitational action is also a generalized version (GSTG), in which different connections are supposed to yield different cosmic evolution. In this action, since the non-metricity scalar $(Q)$ appears linearly, so all the problems associated with perturbative analysis disappear. This is a major issue. Nonetheless, it takes the same form as scalar-tensor theory of gravity which has been meticulously studied over several decades.\\

Succinctly, curvature induced inflation and late-time accelerated expansion along with a viable radiation and early matter-dominated era are the special and attractive features of modified theory of gravity ($f(R)$ theory), which is absent from teleparallel gravity theories. In a nutshell, the present analysis in the FLRW background with a particular for of $f(X) = \alpha X - \beta X^2$ in association of a minimally coupled scalar field exhibits following facts: i) The theory is free from the pathology of Branched Hamiltonian. ii) early inflation requires to be driven by a scalar field, in which case, the inflationary parameters agree well with observation for flat potential, giving way to the hot Big-Bang upon reheating iii) A viable radiation era is not admissible without a scalar field, however small. iii) Considering the flat potential which agrees best with the observed inflationary parameters, the scalar field escalates exorbitantly in the radiation dominated era and remains constant in the early matter dominated era. iv) Late-time cosmic acceleration cannot be driven from purely geometric teleparallel gravity. v) The covariant scalar-tensor formalism of $f(\mathrm{T})$ theory and scalar-vector-tensor formalism of $f(Q)$ theory can serve the purpose, but the scalar field is retained throughout the cosmic evolution in the form of dark energy.\\ 

 \section*{Author Contributions}
A.K. Sanyal proposed the problem, D. Saha and J.P. Saha carried out the formulation and A.K. Sanyal verified it. D. Saha prepared the draft primarily, and J.P. Saha modified and edited the primary draft. The final version was prepared by A.K. Sanyal. All authors have read and agreed to the final version of the manuscript.

\section*{Funding Statement}
This research received no specific grant from any funding agency in public, commercial or not-for profit sectors.
\section*{Conflict of Interest}

All the authors declare that they have no conflict of interest related to this research. We have no personal or financial relationships that could influence our work.
\section*{Competing Interest}
All the authors declare that they have no significant competing financial, professional or personal interest that might have influenced the performance or presentation of the work described in this manuscript.

\section*{Data Availability}
The authors confirm that the data supporting the findings of this study are available within the article [and/or] its supplementary materials.

\end{document}